\documentclass[11pt]{article}
\pdfoutput=1

\usepackage{cancel,slashed}
\usepackage{bbm}
\usepackage{bm}
\usepackage{amsmath,amsfonts,amssymb,mathtools,nicefrac,nccmath,cases}
\usepackage{graphicx}
\usepackage{cite}
\usepackage{skak}

\usepackage{dsfont}
\usepackage{latexsym}
\usepackage{color}
\usepackage[hyperfootnotes=false,linktocpage]{hyperref}
\usepackage{transparent}
\usepackage[hang,flushmargin]{footmisc}
\usepackage{blkarray}
\usepackage{multirow}
\usepackage{empheq}
\usepackage{comment}
\usepackage{wasysym}
\usepackage{tikzsymbols}

\usepackage{caption}
\usepackage{subcaption}
\usepackage[normalem]{ulem}

\newcommand{\bmat}{\left(\begin{array}}
\newcommand{\emat}{\end{array}\right)}

\def\R{\mathbbm{R}}

\def\a {\alpha}
\def\b {\beta}

\def\K{\mathbf{K}}

\def\1{{\bf 1}}
\def\2{{\bf 2}}
\def\3{{\bf 3}}
\def\4{{\bf 4}}
\def\6{{\bf 6}}

\def\targ#1#2{\genfrac{[}{]}{0pt}{}{#1}{#2}}
\def\targ2#1#2{\genfrac{}{}{0pt}{}{#1}{#2}}

\definecolor{mygr}{rgb}{0,0.6,0}
\definecolor{mygrey}{rgb}{0,0.1,0.2}
\definecolor{myblue}{rgb}{0,0.5,0.9}
\definecolor{myblue2}{rgb}{0,0.5,0.5}
\definecolor{myblue3}{rgb}{0,0.7,0.9}
\definecolor{myblue4}{rgb}{0,0.6,0.6}
\definecolor{myorange}{rgb}{1,0.5,0}
\definecolor{mypurple}{rgb}{0.6,0,1}
\definecolor{mygolden}{rgb}{1,0.8,0.2}
\definecolor{mycyan}{rgb}{0,1,1}
\definecolor{mymagenta}{rgb}{1,0,1}
\definecolor{mykiwi}{rgb}{0.8,1,0.5}
\definecolor{mybrown}{cmyk}{0.14, 0.42, 0.56, 0.2}
\definecolor{myturq}{cmyk}{0.99, 0, 0.2, 0.4}
\definecolor{myaubergine2}{cmyk}{0.4, 0.5, 0, 0.1}
\definecolor{myaubergine}{cmyk}{0.6,0.85,0,0}
\definecolor{CycleGreen}{cmyk}{0.52,0,1,0}
\definecolor{CycleBrown}{cmyk}{0, 0.4, 0.9, 0.2}

\DeclareFontFamily{U}{rcjhbltx}{}
\DeclareFontShape{U}{rcjhbltx}{m}{n}{<->rcjhbltx}{}
\DeclareSymbolFont{hebrewletters}{U}{rcjhbltx}{m}{n}

\DeclareMathSymbol{\lamed}{\mathord}{hebrewletters}{108}
\DeclareMathSymbol{\mem}{\mathord}{hebrewletters}{109}
\DeclareMathSymbol{\ayin}{\mathord}{hebrewletters}{96}
\DeclareMathSymbol{\tsadi}{\mathord}{hebrewletters}{118}
\DeclareMathSymbol{\qof}{\mathord}{hebrewletters}{113}
\DeclareMathSymbol{\resh}{\mathord}{hebrewletters}{114}
\DeclareMathSymbol{\pe}{\mathord}{hebrewletters}{112}
\DeclareMathSymbol{\pesofit}{\mathord}{hebrewletters}{80}
\DeclareMathSymbol{\samekh}{\mathord}{hebrewletters}{115}
\DeclareMathSymbol{\tav}{\mathord}{hebrewletters}{116}
\DeclareMathSymbol{\vav}{\mathord}{hebrewletters}{119}
\DeclareMathSymbol{\het}{\mathord}{hebrewletters}{120}
\DeclareMathSymbol{\yod}{\mathord}{hebrewletters}{121}
\DeclareMathSymbol{\zayin}{\mathord}{hebrewletters}{122}
\DeclareMathSymbol{\alephdot}{\mathord}{hebrewletters}{128}
\DeclareMathSymbol{\tsadisofit}{\mathord}{hebrewletters}{90}
\DeclareMathSymbol{\shin}{\mathord}{hebrewletters}{152}

\newtheorem{conjecture}{Conjecture}

\def\CN {{\cal N}}
\def\CM {{\cal M}}
\def\CK {{\cal K}}
\def\CV {{\cal V}}
\def\CY {{\cal Y}}

\def\sig{{\sigma}}
\def\del{{\delta}}

\def\be{\begin{equation}}
\def\ee{\end{equation}}
\def\bea{\begin{eqnarray}}
\def\eea{\end{eqnarray}}
\def\bes{\begin{subequations}}
\def\ees{\end{subequations}}

\def\oh{\frac{1}{2}}

\def\im{\mbox{Im}\, }

\def\om{\omega}

\usepackage{multicol}
\usepackage{float}
\usepackage{caption}
\def\p {{\partial}}

\def\g {{\gamma}}

\def\CO {{\cal O}}

\newcommand{\cF}{\mathcal{F}}

\newcommand{\cK}{\mathcal{K}}
\newcommand{\cM}{\mathcal{M}}

\newcommand{\cO}{\mathcal{O}}

\newcommand{\cT}{\mathcal{T}}
\newcommand{\cI}{\mathcal{I}}

\newcommand{\IR}{\mathbb{R}}

\newenvironment{eqn}{\begin{equation}\begin{aligned}}{\end{aligned}\end{equation}\noindent}
\newenvironment{eqn*}{\begin{equation*}\begin{aligned}}{\end{aligned}\end{equation*}\noindent}

\makeatletter
\newsavebox\myboxA
\newsavebox\myboxB
\newlength\mylenA

\newcommand*\xoverline[2][0.75]{%
\sbox{\myboxA}{$\m@th#2$}%
\setbox\myboxB\null
\ht\myboxB=\ht\myboxA%
\dp\myboxB=\dp\myboxA%
\wd\myboxB=#1\wd\myboxA
\sbox\myboxB{$\m@th\overline{\copy\myboxB}$}
\setlength\mylenA{\the\wd\myboxA}
\addtolength\mylenA{-\the\wd\myboxB}%
\ifdim\wd\myboxB<\wd\myboxA%
   \rlap{\hskip 0.5\mylenA\usebox\myboxB}{\usebox\myboxA}%
\else
    \hskip -0.5\mylenA\rlap{\usebox\myboxA}{\hskip 0.5\mylenA\usebox\myboxB}%
\fi}
\makeatother

\begin{document}
\pagestyle{plain}

\makeatletter
\@addtoreset{equation}{section}
\makeatother
\renewcommand{\theequation}{\thesection.\arabic{equation}}

\pagestyle{empty}
\rightline{IFT-UAM/CSIC-23-144}
\vspace{0.5cm}
\begin{center}
\Huge{{On the moduli space curvature at infinity} 
\\[10mm]}
\normalsize{Fernando Marchesano,$^{1}$ Luca Melotti\,$^{1,2}$ and Lorenzo Paoloni\,$^{1}$}\\[12mm]
\small{
${}^{1}$ Instituto de F\'{\i}sica Te\'orica UAM-CSIC, c/ Nicol\'as Cabrera 13-15, 28049 Madrid, Spain \\[2mm] 
${}^{2}$ Departamento de F\'{\i}sica Te\'orica, Universidad Aut\'onoma de Madrid, 28049 Madrid, Spain
\\[10mm]} 
\small{\bf Abstract} \\[5mm]
\end{center}
\begin{center}
\begin{minipage}[h]{15.0cm}

We analyse the scalar curvature of the vector multiplet moduli space $\CM^{\rm VM}_X$ of type IIA string theory compactified on a Calabi--Yau manifold $X$. While the volume of $\CM^{\rm VM}_X$ is known to be finite, cases have been found where the scalar curvature diverges positively along trajectories of infinite distance. We classify the asymptotic behaviour of the scalar curvature for all large volume limits within $\CM^{\rm VM}_X$, for any choice of $X$, and provide the source of the divergence both in geometric and physical terms. Geometrically, there are effective divisors whose volumes do not vary along the limit. Physically, the EFT subsector associated to such divisors is decoupled from gravity along the limit, and defines a rigid $\CN=2$ field theory with a non-vanishing moduli space curvature $R_{\rm rigid}$. We propose that the relation between scalar curvature divergences and field theories decoupled from gravity is a common trait of moduli spaces compatible with quantum gravity.

\end{minipage}
\end{center}
\newpage
\setcounter{page}{1}
\pagestyle{plain}
\renewcommand{\thefootnote}{\arabic{footnote}}
\setcounter{footnote}{0}


\tableofcontents


\section{Introduction}
\label{s:intro}

The Swampland Programme \cite{Vafa:2005ui,Brennan:2017rbf,Palti:2019pca,vanBeest:2021lhn,Grana:2021zvf} aims to describe the set of effective field theories (EFTs) with a gravitational UV completion. One way to do so is to establish general constraints on the range value and the field dependence for the couplings of such EFTs. This was the spirit behind \cite{Ooguri:2006in}, where several criteria were proposed to constrain the geometry of the moduli space of vacua $\cM$ of any EFT compatible with Quantum Gravity. One of these proposals conjectured that $\cM$ should always have points at infinite field distance, and another one that the scalar curvature of $\cM$ near such points should be non-positive. Part of the intuition came from compactifications of type II string theory on Calabi--Yau (CY) manifolds, which lead to 4d $\CN=2$ supergravity EFTs. In particular, the vector multiplet moduli space of such theories $\CM_{\rm 4d}^{\rm VM}$ is endowed with a Weil-Petersson (WP) metric that is known to be exact, that always contains points at infinite distance and that provides $\CM_{\rm 4d}^{\rm VM}$ with a finite volume \cite{Todorov:2004jg,Lu:2005bi,Lu:2005bj}. Moreover, for cases with a single modulus $T$ the K\"ahler potential that describes the asymptotic WP metric near the infinite-distance point $\im T = \infty$ reads $K = - 3 \log \im T$ and leads to an asymptotically constant, negative curvature. 

However, it is precisely in the context of type II Calabi--Yau compactifications where counterexamples to the curvature proposal were found \cite{trenner2010asymptotic}. The counterexamples involved 4d $\CN=2$ vector multiplet moduli spaces with three moduli, and the scalar curvature of the WP metric was not only asymptotically positive, but actually divergent. More recently, these examples have been reconsidered in \cite{Cecotti:2021cvv}, where it was pointed out that the divergence disappears if in $\CM_{\rm 4d}^{\rm VM}$ one replaces the WP metric by the Hodge metric. It was then proposed that scalar curvature divergences are related to the lack of completeness of the WP metric, and that the same physical mechanism should shield the EFT from both of these effects.

In this work we aim to shed some light on the asymptotic behaviour of the moduli space scalar curvature. For this, we classify its behaviour in a wide class of infinite distance limits, and identify the physical origin of the divergences whenever they appear. We focus our analysis on infinite-distance limits within the 4d $\CN=2$ vector multiplet moduli space obtained from type IIA string theory compactified on an arbitrary Calabi--Yau three-fold $X$. These have been recently classified in terms of monodromy techniques \cite{Grimm:2018ohb,Grimm:2018cpv,Corvilain:2018lgw} and the Emergent String Conjecture \cite{Lee:2019wij}, with the motivation to understand how the Distance Conjecture (SDC) \cite{Ooguri:2006in}  is realised along them. More precisely, we  focus on limits where the volume of $X$ tends to infinity, which were also considered in \cite{Marchesano:2022axe} in the context of the Emergence Proposal \cite{Harlow:2015lma,Grimm:2018ohb,Heidenreich:2018kpg}. Large-volume limits have the advantage that they can be easily mapped to trajectories in the 5d vector multiplet moduli space $\CM_{\rm 5d}^{\rm VM}$ of M-theory compactified on $X$, where they become either of finite or infinite length. This 5d description is  instrumental to understand which limits give rise to a divergence. 

In general, we obtain that divergences for the type IIA scalar curvature $R_{\rm IIA}$ are related to divisors ${\cal D} \subset X$ whose volume stays constant along the large-volume limit. From the M-theory perspective it is the volume of $X$ that stays constant along $\CM_{\rm 5d}^{\rm VM}$, while ${\cal D}$ shrinks to zero size. In particular, if the trajectory becomes of finite distance in  $\CM_{\rm 5d}^{\rm VM}$, then ${\cal D}$ must be a generalised del Pezzo divisor that shrinks to a point. Shrinkable cycles are known to give rise to limits in which a non-trivial field theory decouples from gravity \cite{{Kachru:1995fv,Klemm:1996bj}}. We argue that this is also what happens along the type IIA infinite-distance trajectories where $R_{\rm IIA}$ diverges. 

The Distance Conjecture predicts that, as we approach an infinite distance point of $\CM$, the maximal cut-off of an EFT $m_*$ decreases to zero in Planck units. If along such a limit there is a field theory sector that remains dynamical below $m_*$, then it will be gradually decoupled from gravity, resulting in a rigid field theory that dominates the low-energy dynamics. In our type IIA setup $m_*$ is the D0-brane mass, and the constant-volume divisor ${\cal D}$ hosts a 4d $\CN=2$ field theory subsector that remains dynamical below this scale. The moduli space and kinetic terms of this subsector is described by a rigid special K\"ahler geometry \cite{Freed:1997dp}, that it is recovered from the local special K\"ahler structure that describes  $\CM_{\rm 4d}^{\rm VM}$, as in other gravity-decoupling limits \cite{Seiberg:1994rs,Andrianopoli:1996cm,Katz:1996fh,Lerche:1996ni,Katz:1997eq,Billo:1998yr,Gunara:2013rca}. Using the standard expressions for the scalar curvature in both special K\"ahler geometries, one can see that a non-trivial curvature $R_{\rm rigid}$ for the $\CN=2$ rigid field theory below $m_*$ results in a (positive) divergence for the $\CN=2$ supergravity curvature measured in Planck units, that goes like $R_{\rm IIA} \simeq R_{\rm rigid} \frac{M_{\rm P}^2}{m_*^2}$. Via Conjecture \ref{conj:CC} we propose that this is a general mechanism valid beyond our specific setup, and that it essentially accounts for all the divergences that a moduli space scalar curvature can have in theories that are compatible with quantum gravity. 

For those limits along which the curvature does not diverge there is a variety of behaviours, which we classify for the type IIA EFT string limits of \cite{Marchesano:2022axe}. We argue that this is a representative subset of limits, in the sense that it is dense in the continuum of type IIA large-volume endpoints, a set that can be identified with the moduli space $\CM_{\rm 5d}^{\rm VM}$. We find that non-divergent scalar curvatures are asymptotically negative, unless we consider a limit that it is close to a divergent one in $\CM_{\rm 5d}^{\rm VM}$. As a result, limits with constant positive curvature display a hierarchy of gauge couplings, and in particular some gauge sectors that dominate over the remaining gauge interactions and the gravitational ones, although not parametrically like in the divergent case. If this was a general lesson, one would recover the physical intuition behind the scalar curvature proposal in \cite{Ooguri:2006in}, in the sense that whenever gravity is not substantially suppressed with respect to other interactions one recovers an asymptotic scalar curvature that is negative.  

The rest of this paper is organised as follows. Section \ref{s:typeIIA} describes the type IIA infinite-distance limits that we consider and their M-theory description. Section \ref{s:curvature} describes the moduli space scalar curvature along such limits, and identifies its divergences both from an M-theory and a 4d EFT perspective, leading to Conjecture \ref{conj:CC}. Section \ref{s:limits} analyses the asymptotic behaviour of the curvature along different classes of EFT string limits, which is summarised in table \ref{tab:curv(w,rk)}. Section \ref{s:examples} contains some examples that illustrate the general picture, and in section \ref{s:conclu} we discuss some possible implications of our findings. Some technical details have been relegated to the appendices. Appendix \ref{ap:details} contains the perturbative expansion of the asymptotic metric and appendix \ref{ap:degenerate} discusses the degeneracy of the different EFT string limits. Both sets of results are used in the classification of limits performed in section \ref{s:limits}. 


\section{Type IIA large volume limits}
\label{s:typeIIA}

Let us consider type IIA string theory compactified on a Calabi--Yau three-fold $X$ 
\be
ds^2 = ds_{\R^{1,3}}^2 + ds_{X}^2 \, ,
\label{comp}
\ee
whose metric is determined by the periods of its holomorphic three-form $\Omega$ and its K\"ahler form $J$. The latter can be expanded as
\be
J = t^a \omega_a\, , \qquad a =1, \dots h^{1,1}(X) ,
\ee
where $\ell_s^{-2} \omega_a$ is a basis of integral harmonic two-forms Poincar\'e dual to a basis of Nef divisors, with $\ell_s = 2\pi \sqrt{\a'}$ the string length. We use the same basis to  reduce the B-field and RR potentials
\be
B = b^a \omega_a + \dots\, , \qquad C_1 = \kappa_4 A_1^0\, , \qquad C_3 = \kappa_4 \hat{A}_1^a \wedge \omega_a + \dots\, ,
\label{BCs}
\ee
where the dots stand for terms of the expansion that are irrelevant for the vector multiplet sector. Here $T^a =  b^a + i t^a$ and $A_1^0$, $\hat{A}_1^a$ represent 4d complex scalars and vector bosons, respectively. Upon dimensional reduction of the standard type IIA 10d supergravity action one recovers the following piece of 4d $\CN=2$ Lagrangian describing the vector multiplet sector \cite{Ferrara:1988ff,Andrianopoli:1996cm,Lauria:2020rhc}
\be
S_{\rm 4d}^{\rm VM} =  \frac{1}{2\kappa_{4}^2} \int_{\R^{1,3}} R * \mathbbm{1} - 2 g_{ab} dT^a \wedge * d\bar{T}^{\bar{b}} +  \oh \int_{\R^{1,3}}  I_{AB} F^A \wedge *_4 F^B + R_{AB} F^A \wedge F^B ,
\label{SVM}
\ee
with $A = (0,a)$. Here $F^A$ are integrally-quantised two-form field strengths associated to the U(1) gauge bosons: $F^0 = dA_1^0$ is the graviphoton field strength, while $F^a = dA_1^a \equiv d(\hat{A}_1^a - b^a A_1^0)$. The moduli space metric in the large-volume regime is specified by
\be
g_{ab} = \frac{3}{2} \left( \frac{3}{2} \frac{\CK_a \CK_b}{\CK^2} - \frac{\CK_{ab}}{\CK} \right)  ,
\label{metric}
\ee
with $\CK_{abc} = \ell_s^{-6} \int \omega_a\wedge \omega_b \wedge \omega_c$ the triple intersection numbers of $X_6$, from where we build $\CK_{ab} = \CK_{abc} t^c$, $\CK_{a} = \CK_{abc} t^bt^c$ and $\CK =  \CK_{abc} t^at^bt^c = 6 {\cal V}_{X}$. The gauge couplings are  
\be
I\, =\, - \frac{\CK}{6}
\left(
\begin{array}{cc}
1 +   4 g_{ab} b^ab^b & 4 g_{ab} b^b \\ 
4 g_{ab} b^b & 4 g_{ab}
\end{array}
\right) , \qquad
R\, =\, -
\left(
\begin{array}{cc}
 \frac{1}{3} \CK_{abc}b^ab^bb^c  & \oh  \CK_{abc}b^bb^c 
 \\  \oh  \CK_{abc}b^bb^c 
 & \CK_{abc}b^c 
\end{array}
\right) . 
\label{IandR}
\ee
These expressions are a large-volume approximation that receive curvature and world-sheet instanton corrections. These can be encoded in terms of the full prepotential 
\be
 {\cal F} = -\frac{1}{6} \cK_{abc}T^aT^bT^c + \oh K_{ab}^{(1)}T^aT^b + K_{a}^{(2)}T^a + \frac{i}{2} K^{(3)} + (2\pi i)^{-3} \sum_{\bm{k}} n_{\bm{k}}^{(0)} {\rm Li}_3 \left( e^{2\pi ik_aT^a} \right) ,
\label{fullF}
\ee
where $K^{(3)} = \frac{\zeta(3)}{8\pi^3} \chi(X_6)$,  $K_{ab}^{(1)}$ and $K_{a}^{(2)}$ depend on topological data of $X$, Li$_3$ stands for the 3rd polylogarithmic function and $n_{\bm{k}}^{(0)}$ for the genus zero Gopakumar-Vafa (GV) invariant of the curve class $k_a {\cal C}^a$, with ${\cal C}^a$ the dual of $[\ell_s^{-2} \om_a]$. In the large-volume regime the piece $\cF^{\rm cl} = -\frac{1}{6} \cK_{abc}T^aT^bT^c$ is the dominant one, and it is from this term that \eqref{metric} and \eqref{IandR} are derived. The effect of the polynomial corrections on these quantities have been discussed in \cite[Appendix B]{Marchesano:2022axe}, and they can be neglected as long as all volumes are large. The same applies to the exponentially-suppressed terms that correspond to world-sheet instanton corrections, although we will see that, when discussing the moduli space curvature, in some cases they may play a role. 

In this setup, \cite{Marchesano:2022axe} explored a particular class of infinite distance trajectories, in which a linear combination of K\"ahler moduli specified by a vector $\bm{e}$ grows towards infinite volume
\begin{equation}
t^a = t^a_0 + e^a \phi, \quad \text{with} \quad \phi \to \infty ,
\label{limita}
\end{equation}
with $t^a_0$ inside the K\"ahler cone and $e^a \in \mathbbm{N}$. Such limits are motivated by the backreaction of 4d strings in the theory, built by wrapping an NS5-brane on the divisor class ${\cal D}_{\bm{e}} = \ell_s^{-2} e^a [\om_a]$. These strings are magnetically charged under the axions $b^a$, with charges $e^a$, and the trajectory \eqref{limita} corresponds to the profile of the saxions $t^a$ as one approaches the string core. Using the terminology of \cite{Lanza:2020qmt,Lanza:2021udy,Lanza:2022zyg}, they were dubbed EFT string limits. Here we are not interested in probing the physics close to an EFT string, rather we take this particular trajectory in the vector multiplet moduli space, inspired by the flow associated to such strings. In order to keep the field trajectory within the vector multiplet moduli space one must rescale the 10d dilaton accordingly
\begin{equation}
g_s (\phi) = g_{s, 0}  \frac{{\cal V}_{X}^{1/2} (\phi)}{{\cal V}_{X,0}^{1/2}} \to \infty ,
\label{codila}
\end{equation}
taking us to a region of 10d strong coupling. Nevertheless, from the 4d EFT perspective we are led to a weakly coupled region, with a simpler microscopic description in terms of a dual frame.

Trajectories of the form \eqref{limita} are only a subset of the whole set of infinite distance limits in the vector multiplet moduli space of type IIA Calabi-Yau compactifications, which have been recently classified in \cite{Corvilain:2018lgw,Lee:2019wij}. In general one obtains three main types of limits, each with a different dual description. Applied to EFT string limits such a classification translates into:

\begin{enumerate}
    \item ${\bf k} \equiv \CK_{abc}e^ae^be^c\neq 0$, dubbed type IV$_d$ or $n=3$ limits in \cite{Corvilain:2018lgw}. Their dual description is M-theory compactified on $S^1 \times X$, where the $S^1$ radius tends to infinity. 
    \item ${\bf k} = 0$, but ${\bf k}_a \equiv \CK_{abc} e^be^c \neq 0$ for some $a$, dubbed type III$_c$ or $n=2$ limits in \cite{Corvilain:2018lgw}, and J-Class A limits in  \cite{Lee:2019wij}. Their dual description consist of F-theory compactified on ${\bf T}
^2  \times X$, with ${\bf T}^2$ taken to a decompactification limit.  
    \item ${\bf k}_a  = 0$, $\forall a$, dubbed type II$_b$ or $n=1$  limits in \cite{Corvilain:2018lgw}, and J-Class B limits in \cite{Lee:2019wij}. Their dual description features a weakly-coupled,  asymptotically tensionless critical string \cite{Lee:2019wij}. 
\end{enumerate}

The subindices in II$_b$, III$_c$, and IV$_d$ correspond to non-negative integers that classify subclasses of limits, and which for EFT string limits are determined by 
\be
r= {\rm rank}\, \K , \qquad \K_{bc} \equiv e^a \CK_{abc} \, .
\label{rank}
\ee
In particular, it follows from the analysis in \cite{Corvilain:2018lgw} that $b=d =r$, while  $c =r-2$. A physical interpretation for the subindex in II$_b$ was given in \cite{Marchesano:2022axe}, in terms of towers of asymptotically massless states. In the following sections we will see that this lesson is more general and all the subindices are crucial to describe the set of towers of asymptotically massless states for this kind of limits, as well as the asymptotic behaviour of the moduli space scalar curvature. 

Finally, the above classification of limits has a physical interpretation in terms of 4d EFT string solutions \cite{Lanza:2020qmt,Lanza:2021udy,Lanza:2022zyg}. In this setup, the relevant EFT strings are made up of NS5-brane wrapping Nef divisors $e^aD_a$ dual to $\ell_s^{-2}e^a[\omega_a]$ in $X_6$, and whose 4d backreaction generates the saxion flow \eqref{limita} towards their core. The 4d tension ${\cal T}$ of an EFT string varies as $\phi^{-1}$ along its own flow, and sets a reference mass scale $\sqrt{\cal T}$ that can be compared with the scale $m_*$ of the leading tower of asymptotically massless states. It was proposed in \cite{Lanza:2021udy,Lanza:2022zyg} that along 4d EFT string limits this relation is universal and quite simple, namely of the form
\begin{equation}
    \frac{m_*^2}{M_{\rm P}^2} \sim \left(\frac{\cal T}{M_{\rm P}^2} \right)^w , \qquad w=1,2,3 \, ,
    \label{scaling}
\end{equation}
where $w$ is dubbed the scaling index. In the setup at hand, one can easily check that $m_* \simeq m_{\rm D0}$ is specified by the mass of a D0-brane and that $w=n$ corresponds to the singularity type of each limit, as defined in \cite{Grimm:2018ohb,Grimm:2018cpv,Corvilain:2018lgw}.

\subsection*{The M-theory perspective}

As stressed in \cite{Lee:2019wij}, type IIA limits with scaling index $w=3$ have a simple interpretation in terms of M-theory compactified on the Calabi--Yau $X$  \cite{Cadavid:1995bk,Ferrara:1996hh,Ferrara:1996wv}, as they correspond to a particular point in the interior of the resulting 5d vector multiplet moduli space. To see this, let us translate the type IIA trajectory \eqref{limita} in terms of M-theory compactified on $S^1 \times X$, assuming for simplicity a vanishing vev $b^a =0$ for the B-field axions. From a 11d perspective, the flow of moduli reads
\be
2\pi R_{S^1} (\phi) =   \frac{ \CV_X^{1/3}}{\CV_M^{1/3}} , \qquad t^a_M (\phi) = t^a \frac{ \CV_M^{1/3}}{\CV_X^{1/3}} ,
\label{flow11d}
\ee
with $R_{S^1}$ the $S^1$ radius and $t^a_M$ the K\"ahler moduli of $X$, both measured in 11d Planck units $\ell_{11}$. Here $ \CV_M = g_s^{-2} \CV_X$ stands for the volume of $X$ in units of $\ell_{11}$, a quantity that remains constant along the trajectory. Then, if the $S^1$ radius is much larger than the internal sizes of $X$, it makes sense to describe the trajectory from a 5d perspective:
\be
2\pi R_{5} (\phi) =   \CV_X^{1/3}  , \qquad
M^a(\phi)  = \frac{t^a}{ \CV_X^{1/3}} ,
\label{flow5d}
\ee
where $R_{5}$ is the $S^1$ radius measured in 5d Planck units $\ell_5 = \ell_{11} \CV_M^{-1/3}$. The set $\{M^a\}$ is nothing but the K\"ahler coordinates of $X$ normalised such that $\CK_{abc} M^a M^b M^c =6$, which is a necessary constraint to stay in the 5d vector multiplet moduli space $\CM_{\rm 5d}^{\rm VM}$ of M-theory compactified on $X$. To parametrise $\CM_{\rm 5d}^{\rm VM}$ one may either drop this normalisation and see $\CM_{\rm 5d}^{\rm VM}$ as a set of real projective classes $[\bm{t}_M] = \{ \lambda \bm{t}_M | \lambda \in \IR\}$, or instead define a set of affine coordinates $\psi^i$, with $i = 1, \dots , h^{1,1}(X)-1$. With  this second option, the bosonic action of this sector reads
\begin{align}\nonumber
S_{\rm 5d}^{\rm VM} & =  \frac{2\pi}{\ell_5^3} \int_{\R^{1,4}} \left(R * \mathbbm{1} - \frac{1}{2} \mathfrak{g}_{ij} d\psi^i \wedge * d\psi^j\right)  - \frac{1}{4\pi \ell_5}   \int_{\R^{1,4}} {\cal I}_{ab} F^a \wedge *_5 F^b \\ & -  \frac{1}{24\pi^2}  \int_{\R^{1,4}} \CK_{abc}\, A^a \wedge F^b \wedge F^c .
\label{SVM5d}
\end{align}
Of particular interest to us will be the 5d gauge kinetic matrix ${\cal I}_{ab}$, which has a simple expression in terms of the variables $M^a$
\be
\cI_{ab} = \frac{1}{4} \CK_{a}^M \CK_{b}^M  - \CK_{ab}^M  , 
\label{cali}
\ee
where $\CK^M_a = \CK_{abc}M^bM^c$ and $\CK^M_{ab} = \CK_{abc}M^c$. Notice that $\cI_{ab}$ is nothing but the image of the 4d kinetic matrix $I_{ab}$ in \eqref{IandR} under the map  $t^a \mapsto M^a = t^a/\CV_X^{1/3}$. Finally, the kinetic matrix for the scalars can be expressed as $\mathfrak{g}_{ij} = \cI_{ab} \p_i M^a \p_j M^b$.  

Asymptotically, the $S^1$ radius behaves as $R_{S^1} \sim \phi^{w/3}$, and so we are naively led to a decompactification limit to 5d. This is clearly the case for $w=3$ limits, because then ${\bf k} \neq 0$ and the K\"ahler trajectory \eqref{flow5d}  in terms of the variables $M^a$ reads
\be
M^a(\phi) = \left(\frac{6}{{\bf k}}\right)^{1/3}  \left( e^a + \phi^{-1} \left( t_0^a - e^a \frac{{\bf k}_b t_0^b}{{\bf k}}\right)  + \CO(\phi^{-2}) \right) .
\label{Mtrajw3}
\ee
Thus, as $\phi \to \infty$ we reach a point in 5d vector moduli space that is proportional to $\bm{e}$. Since all the components of $M^a (\phi=\infty)$ are finite, this translates into a finite-distance point in 5d vector multiplet moduli space $\CM_{\rm 5d}^{\rm VM}$, as advanced above. If some of the entries of $\bm{e}$ vanish, one is in particular led to a point in 5d moduli space where some contractible curves or divisors in $X$ shrink, while the rest remain of finite volume.\footnote{This is for instance the case for the so-called {\em elementary} EFT string limits \cite{Lanza:2021udy}, that generate the K\"ahler cone of $X$, and where only one component of $\bm{e}$ is non-vanishing and equal to one.\label{ft:elementary}} In this case one lands in a boundary of the K\"ahler cone  $\CK_X$ of $X$, which may either be an actual finite-distance boundary of the vector moduli space of the 5d EFT, or instead a point where a phase transition occurs \cite{Witten:1996qb}. In this second case the moduli space $\CM_{{\rm 5d},X}^{\rm VM}$ is extended by patching, through a flop transition, the K\"ahler cone of a Calabi-Yau manifold $X'$ that is birationally equivalent to $X$. The sum of K\"ahler cones connected by flop transitions ${\CK}_{\rm ext} = \cup_X \CK_X$ is called the extended K\"ahler cone, and it specifies the actual 5d vector multiplet moduli space $\CM_{\rm 5d}^{\rm VM} = \cup_X \CM_{{\rm 5d},X}^{\rm VM}$. Since these loci turn out to be crucial for our analysis, let us stress that the shrinking of cycles only occurs in the M-theory frame. In type IIA units all curves and divisors have large volumes as long as the initial trajectory point $t_0^a$ is in the deep interior of the K\"ahler cone of $X$. This also guarantees that $R_{S^1}^2 \gg t_a^M$ at all points in \eqref{limita}, and it will be our working assumption in the following. 

Things are different for $w=2$ and $w=1$, limits, where ${\bf k} =0$. There, the entries $e^a$ that are non-vanishing correspond to K\"ahler variables $t^a_M$ or $M^a$ that tend to infinity as $\phi\to\infty$, while the vanishing ones correspond to shrinking cycles. Following the discussion in \cite{Lee:2019wij} it is easy to see that $w=2$ trajectories translate into J-Class A limits in the vector moduli space of M-theory, that eventually lead to a further decompactification  to F-theory. Similarly, $w=1$ EFT string trajectories are J-Class B M-theory limits that result in emergent string limits. Even if in both cases the simplest microscopic description is not in terms of M-theory, one can still use the dictionary \eqref{flow5d} to translate the type IIA trajectory \eqref{limita} into an excursion in M-theory moduli space. 
In fact, by describing the 5d vector moduli space as a projective space $\CM_{\rm 5d}^{\rm VM} = \{[\bm{t}_M]\}$, one can map each 4d EFT string charge to a point in $\CM_{\rm 5d}^{\rm VM}$ via $\bm{e} \mapsto [\bm{e}]$. In this correspondence, EFT string charges $\mathbf{e}$ that implement $w=3$ limits are mapped to finite-distance points in $\CM_{\rm 5d}^{\rm VM}$, including finite-distance boundaries. Charges that generate $w=2$ or $w=1$ limits are instead mapped to points in those boundaries of $\CM_{\rm 5d}^{\rm VM}$ that lie at infinite distance in the 5d metric. 

Incidentally, this M-theory perspective gives a specific realisation of the Distant Axionic String Conjecture \cite{Lanza:2020qmt,Lanza:2021udy,Lanza:2022zyg}. Using it one may identify  $\CM_{{\rm 5d},X}^{\rm VM}$, the 5d vector multiplet moduli space of M-theory on $X$, with the endpoints of large-volume infinite-distance limits in $\CM_{{\rm 4d},X}^{\rm VM}$, the 4d vector moduli space of type IIA on $X$. The 4d EFT string flow endpoints are a dense discrete subset in $\CM_{{\rm 5d},X}^{\rm VM}$, described by $[\bm{t}_M] = [\bm{e}]$. Hence, by appropriately choosing an EFT string charge one may get arbitrarily close to any large-volume infinite-distance limit endpoint. 

In the next section we discuss how this dictionary is relevant to describe the asymptotic behaviour of the 4d scalar curvature along type IIA large-volume limits. As it turns out, in this regime the 4d scalar curvature has a very simple expression in terms of 5d variables.


\section{The scalar curvature at infinity}
\label{s:curvature}

In this section we study the scalar curvature $R_{\rm IIA}$ of the vector moduli space of the type IIA compactification  \eqref{comp}, and in particular its dependence along the limits \eqref{limita}. In the  large-volume regime the 4d vector multiplet moduli space metric can be approximated by \eqref{metric}, a tensor that also determines the 5d vector multiplet Lagrangian of M-theory compactified in the same Calabi--Yau, see \eqref{SVM5d}. In fact, at large volumes $R_{\rm IIA}$ has a natural expression in terms of M-theory quantities, a result that we will exploit to classify its asymptotic behaviour along different limits. 

In complex coordinates $(T^a, \bar{T}^{\bar{a}})$, the Riemann curvature associated to the metric \eqref{metric} reads
\be
R_{a\bar{b}c\bar{d}}^{\rm cl} = g_{ab} g_{cd} + g_{ad} g_{bc} - \frac{9}{16\CK^2} g^{ef}  \CK_{ace} \CK_{bdf} \, ,
\label{Riemann}
\ee
where here and below we are expressing the complex-coordinate components on the left hand side in terms of real coordinate tensors on the right hand side. The previous expression follows from the general curvature formula of (local) special K\"ahler manifolds  \cite{Strominger:1990pd,Andrianopoli:1996cm}, applied to a region with axionic shift symmetries like the one at hand  \cite{trenner2010asymptotic,trenner2010curvature}. In particular, we are using the large-volume approximation to replace $\p_a\p_b\p_c\cF \to \p_a\p_b\p_c\cF^{\rm cl} = -\cK_{abc}$.\footnote{This may lead to wrong curvature estimates when some triple intersection numbers vanish, see section \ref{ss:Ftheory}.} From here one deduces that the Ricci tensor is given by
\be
R_{a\bar{b}}^{\rm cl} = - g^{c\bar{d}} R_{a\bar{b}c\bar{d}}^{\rm cl} = - g_{ab} (n_V + 1) +  \frac{9}{16\CK^2} g^{cd} g^{ef}  \CK_{ace} \CK_{bdf} \, ,
\label{Ricci}
\ee
and finally the scalar curvature reads
\be
R_{\rm IIA}^{\rm cl}/2 =  - n_V(n_V + 1) + {\cal Y}^2 \, ,
\label{scalarIIA}
\ee
where $n_V = h^{1,1}(X)$ is the number of vector multiplets and 
\be
{\cal Y }^2 =  \frac{9}{16\CK^2} g^{ab}g^{cd}g^{ef}  \CK_{ace} \CK_{bdf} \, .
\label{Y2IIA}
\ee
That is, we have a negative constant piece plus a moduli-dependent positive piece ${\cal Y }^2$. This implies that in the large-volume region the scalar curvature is bounded from below, while it does not need to be bounded from above and it may even diverge positively. This divergent behaviour was indeed observed in \cite{trenner2010asymptotic} by using a specific expression for the scalar curvature valid for the case $n_V =3$, and then evaluating it in a particular asymptotic region of K\"ahler moduli space. Since such divergences are the current source of counterexamples to Conjecture 3 of \cite{Ooguri:2006in}, in the following we would like to extend the analysis for general Calabi--Yau manifolds and arbitrary large-volume limits. Our  strategy will be to interpret ${\cal Y }^2$ in terms of physical quantities, in order to identify the origin of the divergences. 

In an $\CN=1$ setting one could interpret \eqref{Y2IIA} in terms of physical Yukawa couplings, while the analogous $\CN=2$ quantities are dubbed Pauli couplings and are related to the anomalous magnetic moment of the gauginos. To describe their magnitude, it is natural to consider the axion-independent gauge kinetic terms $\tilde{I}_{AB} \equiv - I_{AB}|_{b^a=0}$ considered in \cite{Marchesano:2022axe}, which read
\be
\tilde{I}_{00}  = \frac{1}{6} \CK\, ,\qquad \tilde{I}_{ab} = \frac{2}{3} \CK g_{ab} \, ,
\label{tildeI}
\ee
with all the remaining components vanishing. This tensor can either be thought of as the gauge coupling of rotated gauge field strengths as in \cite{Marchesano:2022axe}, or an approximation of the gauge couplings in the large volume limit, since $I^{ab} = - \tilde{I}^{ab} -  b^ab^b /\CV_X$ and so the axion dependence disappears at large volume. In any case, in terms of this tensor we find
\be
{\cal Y }^2 =  \CV_X  \tilde{I}^{ab}\tilde{I}^{cd}\tilde{I}^{ef}  \CK_{ace} \CK_{bdf} \, ,
\label{Y2IIAtil}
\ee
which amounts to the triple intersection numbers contracted with the gauge kinetic terms, and multiplied by a volume factor. Recall that in the limits under consideration $\CV_X \sim \phi^w$ and so the volume factor is always a source of divergence. However, we are also lead to a weakly-coupled regime in which the  components $\tilde{I}^{aa} \simeq g_a^2$ typically asymptote as $\phi^{-1}$ or $\phi^{-2}$ \cite{Grimm:2018ohb,Marchesano:2022axe}. Therefore, depending on the limit there may be or not a compensating effect that kills the divergence. 

One way to generate a divergence for ${\cal Y}^2$ is to consider a setup in which a certain gauge coupling $g_{\sig}$ remains constant along \eqref{limita}, and at the same time $\CK_{\sig\sig\sig} \neq 0$. Geometrically, we are asking for the existence of a divisor that remains of constant volume while $\CV_X \to \infty$, or equivalently the existence of a shrinkable divisor in $X$. Such divisors, more precisely generalised del Pezzo surfaces, appear in Calabi--Yau compactifications where after a gravity-decoupling limit one still recovers a non-trivial field theory at low energies \cite{Lerche:1996ni}. In 4d $\CN=2$ vector multiplet settings,  $M_{\rm P} \to \infty$ limits transform a local special K\"ahler geometry into a rigid special K\"ahler one \cite{Andrianopoli:1996cm,Freed:1997dp}, in which (roughly) the K\"ahler potential is no longer of the form $K = - \log  \CK$, but instead $K =\frac{2}{3} \CK$, and the scalar curvature is positive definite and of the form ${\cal Y}^2/\CV_X$. All this suggests that, in moduli space regions where $R_{\rm IIA}^{\rm cl} \simeq 2 {\cal Y}^2$, there is a sector that can be decoupled from gravity and that is dominating  the contributions to the moduli space curvature. 

The remainder of this section aims to make this intuition more precise and to extract its consequences. As we will see, one can arrive to the same conclusion by using two alternative pictures: the 5d M-theory viewpoint and the 4d EFT perspective. 

\subsection{M-theory analysis}
\label{ss:Mth}

In section \ref{s:limits} we analyse the scalar curvature along large-volume limits using \eqref{Y2IIAtil}. However, there is an alternative expression that allows us to build a simple overall picture for the asymptotic behaviour of $R_{\rm IIA}$. Indeed, notice that in terms of M-theory variables $M^a$ we have that
\be
{\cal Y }^2 =  \cI^{ab} \cI^{cd} \cI^{ef}  \CK_{ace} \CK_{bdf} \, ,
\label{Y2M}
\ee
where $\cI^{ab}$ is the inverse of \eqref{cali}. That is, the 4d type IIA moduli space curvature at large volume is a simple function of the gauge kinetic functions of the 5d M-theory compactification on $X$, and more precisely of the physical 5d Chern-Simons  couplings. Moreover, as explained in the last section there is an injective map between the set of EFT string limits \eqref{limita} and the 5d vector moduli space seen as a projective space, via $\bm{e} \mapsto [\bm{e}] \in \CM_{\rm 5d}^{\rm VM}$. This implies that, in order to find out along which limits the IIA curvature tends to infinity one only needs to find out those loci in $\CM_{\rm 5d}^{\rm VM}$ where ${\cal Y }^2$ diverges. It is easy to see that this can only occur at loci where the gauge kinetic matrix $\cI_{ab}$ degenerates. Being a continuous physical quantity that describes 5d gauge couplings, this can only happen at the boundaries of $\CM_{\rm 5d}^{\rm VM}$. 
 
 Indeed, let us consider M-theory compactified on $X$. From here we can construct a 5d vector multiplet  moduli space $\CM_{{\rm 5d},X}^{\rm VM}$ that comes either from {\it i)} taking the projective quotient of the K\"ahler cone $\CK_X$ under the identification $\bm{t}_M \sim \lambda \bm{t}_M$, or {\it ii)} the set of coordinates $\{M^a\}$ where the projective equivalence is fixed such that $\CK_{abc} M^a M^b M^c =6$. The boundaries of $\CM_{{\rm 5d},X}^{\rm VM}$ were classified in \cite{Witten:1996qb} in terms of the facets of $\CK_X$, and they have been recently revisited in the context of the Swampland Programme in \cite{Brodie:2021ain,Alim:2021vhs,Gendler:2022ztv}. One finds four different kinds of boundaries:
 
\begin{itemize}

\item[(i)] A locus where a curve ${\cal C}_a \subset X$ collapses to a point, but no effective divisor collapses. 

This always occurs at finite distance in moduli space, and it is not an actual boundary of the full vector moduli space $\CM_{{\rm 5d}}^{\rm VM}$. Instead, one must adjoin a new K\"ahler cone $\CK_{X'}$ of a birationally equivalent Calabi--Yau $X'$ along this facet, via a flop transition where the triple intersection numbers jump as $\CK_{abc}^\prime = \CK_{abc} - n^0_{{\cal C}_a}  {\cal C}_a {\cal C}_b {\cal C}_c$, with $n^0_{{\cal C}_a}$ the genus zero Gopakumar--Vafa invariant of this class. By adjoining the K\"ahler cones of all birationally equivalent Calabi--Yau manifolds one obtains the extended K\"ahler cone $\CK_{\rm ext} = \cup_X \CK_X$, and from there the actual vector moduli space $\CM_{{\rm 5d}}^{\rm VM} = \cup_X \CM_{{\rm 5d},X}^{\rm VM}$.

At a flop transition a finite number of hypermultiplets become massless, but this does not affect the continuity of the 5d gauge kinetic matrix $\cI_{ab}$, which must remain non-degenerate as the flop occurs. Even if the curvature may suffer a discontinuity due to the jump in triple intersection numbers, ${\cal Y}^2$ must remain finite since $\cI^{ab}$ does not diverge. 

\item[(ii)] A locus where an effective divisor ${\cal D}_\sig \subset X$ collapses to a curve. 

This is an actual boundary of $\CM_{\rm 5d}^{\rm VM}$, that is also reached at finite distance and where a gauge enhancement $U(1) \to SU(2)$ occurs in the vector multiplet sector. The physical origin of such an enhancement is a charged BPS vector multiplet that becomes massless. Again, at this point of enhancement the gauge kinetic matrix $\cI_{ab}$ should remain non-degenerate, which implies that ${\cal Y}^2$ stays finite.

\item[(iii)] A locus where an effective divisor ${\cal D}_\sig \subset X$ collapses to a point. 

This is also a finite-distance boundary of $\CM_{\rm 5d}^{\rm VM}$, that is associated with an infinite tower of BPS particles (M2-branes wrapping curves of ${\cal D}_\sig$) becoming massless, and a magnetic dual BPS string (an M5-brane wrapping ${\cal D}_\sig$) becoming tensionless. An infinite tower of states becoming massless at finite distance in moduli space is only possible together with a strong coupling regime and a non-Lagrangian description of the theory, which in a gravity-decoupling regime becomes a non-trivial 5d SCFT \cite{Seiberg:1996bd,Morrison:1996xf,Douglas:1996xp,Intriligator:1997pq,Aharony:1997ju,Aharony:1997bh}.

The surfaces that realise this possibility are non-Nef effective divisors that are generalised del Pezzo, just as in the setup described above where a divergence for $\CY^2$ was generated. The only difference is that now we are describing the system from an M-theory perspective, so the type IIA finite coupling $g_\sig$ becomes divergent in M-theory units. The divergent 5d coupling and the non-trivial intersection numbers of ${\cal D}_\sig$  signal these boundaries of $\CM_{\rm 5d}^{\rm VM}$ as loci where ${\cal Y}^2$  diverges. Close to these loci ${\cal Y}^2$ is very large, and so \eqref{scalarIIA} is positive. 

\item[(iv)] A boundary at infinite distance.

These are the endpoints of infinite distance limits in $\CM_{{\rm 5d}}^{\rm VM}$, that have been classified in\cite{Lee:2019wij} and recently reconsidered in \cite{Rudelius:2023odg} from the viewpoint of Gopakumar--Vafa invariants. These limits also feature a tower of asymptotically massless particles made up from M2-branes wrapping shrinking curves, and may even include shrinking divisors. However, the physics is very different from the previous case. The tower of states is the one predicted by the SDC, and it is charged under a gauge symmetry with an asymptotically vanishing coupling. Therefore, a priori there is no particular reason to expect a divergence of ${\cal Y}^2$ at these loci. 

From the type IIA viewpoint, these boundaries correspond to either $w=2$ or $w=1$ limits, dual to 6d F-theory compactifications on $X$ and emergent critical strings, respectively. As we will argue in section \ref{s:limits}, in order for $\CY^2$ to diverge asymptotically one needs a divisor ${\cal D}_\sig$ that plays a similar role to generalised del Pezzo in $w=3$ limits.  

\end{itemize} 
 
 To sum up, we find that divergences of the function $\CY^2$ are related to a non-Nef divisor ${\cal D}_\sig$ that shrinks compared to the Nef divisors of the Calabi--Yau. For $w=3$ limits seen from an M-theory perspective the divisor shrinks to a point and the corresponding gauge coupling $g_\sig^M$ tends to infinity, while an infinite tower of particles becomes massless and their dual, non-critical string becomes tensionless. In type IIA string units the divisor ${\cal D}_\sig$ does not shrink, but it is the Nef divisors that tend to infinite volume. Due to the 10d dilaton co-scaling, we also have a tower of states that become massless. They come from D2-branes wrapping curves inside ${\cal D}_\sig$, a D4-brane wrapping ${\cal D}_\sig$ and bound states between them and D0-branes. All these states have a mass that scales like $m_* \simeq m_{\rm D0}$ and therefore, due to the multiplicity of GV invariants, dominate the spectrum of particles below the species scale of the 4d EFT, compared to the KK spectrum of D0-branes. The non-critical string now arises from an NS5-brane wrapping ${\cal D}_\sig$ and, since its tension goes as $\cT_\sig \sim m_*^2$, its oscillation modes must be considered together with the D-brane particle spectrum. Notice that in the IIA frame the gauge coupling $g_\sig$ associated to the shrinkable divisor remains constant along the limit, and that it can be tuned to be arbitrarily weak by appropriately choosing the initial point $\bm{t}_0$ of the trajectory. Therefore, there should not be subtleties related to the stability of the tower of massive charged states. 

\subsection{The 4d EFT perspective}
\label{ss:4dEFT}

The analysis in terms of M-theory data confirms our expectation that those limits where the curvature diverges are related to shrinkable divisors of the Calabi--Yau $X$. Shrinking cycles are also associated to gravity-decoupling limits where, starting from a string compactification, one recovers a gauge field theory without gravity \cite{Kachru:1995fv,Klemm:1996bj}. This procedure has been studied in great detail for the vector multiplet moduli space of type II string theory compactified on a Calabi--Yau \cite{Seiberg:1994rs,Andrianopoli:1996cm,Katz:1996fh,Lerche:1996ni,Katz:1997eq,Billo:1998yr,Gunara:2013rca}, which is precisely our setup. Essentially, by expanding the Calabi--Yau periods in powers of $M_{\rm P}^{-1}$ one recovers the prepotential of a 4d $\CN=2$ rigid supersymmetric gauge theory, whose moduli space is rigid special K\"ahler \cite{Andrianopoli:1996cm}. While one could apply this standard procedure to our setup, in the following we would like to implement a more pedestrian approach, along the lines of \cite{Alexandrov:2017mgi}, combined with the insight of the Distance Conjecture. 

Indeed, let us consider an EFT string limit of the form \eqref{limita}, and consider whether or not there is a non-trivial dynamical  EFT at low energies. A general property of these limits is that, if we restrict the trajectory to the vector multiplet moduli space and so keep $M_s/M_{\rm P}$ constant, the lightest tower of asymptotically massless states predicted by the SDC is the tower of D0-branes, which sets the maximal cut-off scale $m_*$ of a 4d EFT
\be
m_* = \frac{m_{\rm D0}}{\sqrt{8\pi}} = \frac{\sqrt{\pi/2}}{\ell_s g_s} \, .
\ee
where the relative numerical factor is chosen for convenience. If we have a non-trivial 4d field theory along an infinite distance limit, one should be able to express its action in units of $m_*$. Applied to our large-volume EFT string limits we find that the corresponding expression for the 4d kinetic terms of the scalars is
\be
m_*^2 \int_{\R^{1,3}} \tilde{I}_{ab}\, dT^a \wedge * d\bar{T}^{\bar{b}}\, .
\label{rescalekin}
\ee
where the dimensionful prefactor accounts for the fact that the $T^a$ are dimensionless fields. One can absorb this prefactor and the kinetic term  into canonically normalised dimensionful scalar fields. Upon normalisation, the fact that some entries of $\tilde{I}_{ab}$ blow up along the limit $\phi \to \infty$ translates into some scalar fields whose vev  diverge in units of the cut-off $m_*$. Hence, following the discussion in \cite{Gunara:2013rca}, one sees that they should decouple from the dynamics at energies below  $m_*$, where a $\CN=2$ rigid field theory is defined. In terms of \eqref{rescalekin}, one can select which terms should be kept in the rigid theory by doing an expansion of $\tilde{I}_{ab}$ in powers of $\phi$, see eq.\eqref{Kexpand} and Appendix \ref{ap:details}. One deduces that the scalar fluctuations that remain dynamical at the scale $m_*$ are those directions $\bm{t}_{\sig}$ that satisfy 
\be
t^a_\sig {\bf K}_{ab} = 0\,  \quad \forall b\, .
\label{ker}
\ee
That is, field directions that belong to $\ker {\bf K}$. There is an exception to this rule, namely the direction $\bm{t}_{\bm{e}} = \bm{e}$ for a $w=1$ limit, which belongs to $\ker {\bf K}$. Since this field also has a large vev upon canonical normalisation, we do not include it in the rigid theory. Notice that \eqref{rescalekin} recovers the relation  between scalar kinetic terms and gauge kinetic functions that is standard from $\CN=2$ rigid supersymmetry. Accordingly, one can deduce \eqref{ker} in terms of the strength of the gauge interactions. Gauge couplings that belong to $\ker {\bf K}$ stay constant along $\phi \to \infty$, while others vanish and the corresponding gauge interactions become negligible. 

\begin{figure}[ht!]
\begin{center}
\includegraphics[width=14cm]{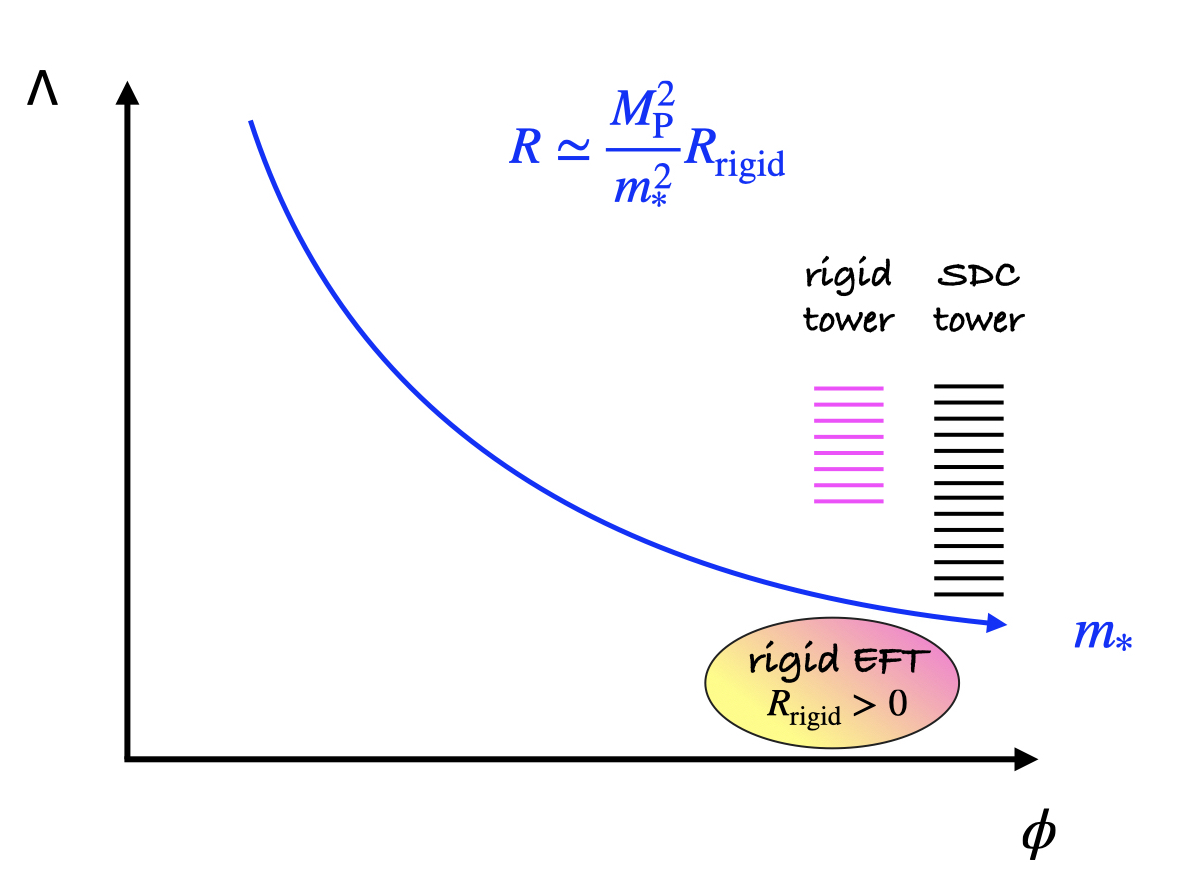}
\caption{Portrayal of a moduli space curvature divergence. Along an infinite distance limit there is a dynamical EFT with non-trivial curvature below the cut-off $m_*$ and so, as a consequence of the SDC, it decouples from gravity. Besides the SDC tower there is a tower charged under the rigid EFT, with similar scaling.  
\label{fig:decoupling}}
\end{center}
\end{figure}

Having selected the set of directions that stay dynamical below $m_*$, one can build the following low energy action
\be
S_{\rm 4d, rigid}^{\rm VM} =m_*^2  \int_{\R^{1,3}} \tilde{I}_{\sig\rho} dT^\sig \wedge * d\bar{T}^{\bar{\rho}} +  \oh \int_{\R^{1,3}}  I_{\sig\rho} F^\sig \wedge *_4 F^\rho + R_{\sig\rho} F^\sig \wedge F^\rho ,
\label{SYM}
\ee
with the tensors $\tilde{I}_{\sigma\rho}$, $I_{\sigma\rho}$, $R_{\sigma\rho}$, restricted to their leading, $\phi$-independent terms with indices along $\ker {\bf K}$. The result indeed corresponds to the vector multiplet sector of the $\CN=2$ field theory obtained from taking the rigid limit. Its classical prepotential is $\cF^{\rm cl}_{\rm rigid} = -\frac{1}{6} \CK_{\sig\rho\tau} T^\sig T^\rho T^\tau + \dots$, where the dots correspond to terms at most linear in the $T^\sig$, that also depend on the vev of the non-dynamical fields $T^\Sigma$ but not on $\phi$. The moduli space scalar curvature is positive and reads
\be
R_{\rm rigid}^{\rm cl} = \frac{1}{2}  \tilde{I}^{\sig\rho}\tilde{I}^{\tau\eta}\tilde{I}^{\mu\nu}  \CK_{\sig\tau\mu} \CK_{\rho\eta\nu} \, .
\label{Rrigid}
\ee
When we look at this curvature in 4d Planck units we obtain
\be
4 \CV_X R_{\rm rigid}^{\rm cl} = 2 \CV_X \tilde{I}^{\sig\rho}\tilde{I}^{\tau\eta}\tilde{I}^{\mu\nu}  \CK_{\sig\tau\mu} \CK_{\rho\eta\nu} \simeq 2 \CY^2 \, ,
\label{divR}
\ee
where in the last term we have used that all the remaining contributions to $\CY^2$ are subleading in powers of $\phi$. Using that \eqref{Rrigid} is $\phi$-independent, we identify the source of curvature divergence along the infinite distance limit $\phi \to \infty$ as a rigid gauge theory that survives in the IR, with a non-trivial scalar curvature, see figure \ref{fig:decoupling}. One can generalise this discussion to understand divergences in the Riemann curvature tensor and sectional curvatures of the moduli space. 

Notice that in terms of the K\"ahler potential $K = - \log \CK$ of the classical metric, condition \eqref{ker} splits the K\"ahler moduli in three sets: $(\bm{t}_0, \bm{t}_{I}, \bm{t}_{II})$, where $\bm{t}_0 = \bm{e}\phi$ is the growing direction, $\bm{t}_{II}$ satisfy \eqref{ker} and $\bm{t}_{I}$ are the remaining moduli. This fits well with the general scheme for rigid limits considered in \cite{Billo:1998yr}, since one has a K\"ahler potential of the form
\be
\kappa_4^2 K = -\log \left(\CK_0  + \CK_{II} + \CK_R \right) ,
\ee
where $\CK_0$ contains all the $\phi$-dependent terms, $\CK_{II}$ all the $\bm{t}_{II}$-dependent terms, and $\CK_R$ the rest. Since $\CK_0 \simeq 6 \CV_X + \CO(\phi^0)$ dominates over the other two terms one can perform the expansion
\be
\kappa_4^2 K = -\log \CK_0 - \frac{\CK_{II}}{\CK_0} + \dots
\ee
In terms of the fields $\bm{t}_{II}$ we obtain a constant term that does not affect the metric, and the following K\"ahler potential for the rigid metric
\be
\kappa_4^2 K_{\rm rigid} = -\frac{1}{6\CV_X} \CK_{II}  \ \to \ m_*^{-2} K_{\rm rigid} =   - \frac{2}{3} \CK_{II} \, .
\ee
where in the last step we have performed a rescaling to measure the metric in units of $m_*$. From here \eqref{Rrigid} follows. 

To connect these results with our discussion of section \ref{ss:Mth} let us analyse the geometric meaning behind the condition \eqref{ker}. Since one can build a basis for $\ker {\bf K}$ made with vectors of integer entries, any element $\bm{u}_\sig \in \ker {\bf K}$ can be interpreted as a divisor class ${\cal D}_\sig$ of $X$. If ${\cal D}_{\bm{e}}$ is the Nef divisor that correspond to the EFT string charge $\bm{e}$, then $\bm{u}_\sig \in \ker {\bf K}$ simply translates into the trivial intersection ${\cal D}_\sig \cdot {\cal D}_{\bm{e}} = 0$. In $w=1$ limits one has one Nef divisor that meets this condition, namely ${\cal D}_{\bm{e}}$, but for any other case ${\cal D}_\sig$ is necessarily non-Nef, because any pair of non-proportional Nef divisors have a non-trivial intersection \cite{Lee:2019wij}. Moreover, whenever ${\cal D}_\sig$ is non-Nef and effective, then it is a shrinkable divisor, because ${\rm Vol}({\cal D}_\sig) = {\cal K}_a v_\sig^a$ is constant along the limit \eqref{limita}, while $\CV_X \sim \phi^w$. Hence, one can perform an overall rescaling of all K\"ahler moduli such that $\CV_X$ stays constant and ${\rm Vol}({\cal D}_\sig) \to 0$. In fact, this is precisely what happens when we map $t^a \mapsto M^a$ and we look at the trajectory from the M-theory perspective of section \ref{s:typeIIA}. 

The M-theory picture is particularly useful for $w=3$ limits, because then the trajectory \eqref{limita} becomes \eqref{Mtrajw3}. In that frame ${\rm Vol}_M({\cal D}_\sig) \to \phi^{-2}$ and the gauge kinetic matrix \eqref{cali} becomes degenerate as $\phi \to \infty$. Both statements indicate that ${\cal D}_A$ is an effective  generalised del Pezzo divisor contracting to a point along this limit, in agreement with the discussion of section \ref{ss:Mth}. Indeed, the fact that $\cI_{ab}$ degenerates shows that we are reaching a boundary of $\CM_{{\rm 5d},X}^{\rm VM}$, but it cannot be a flop transition nor a boundary of $SU(2)$ enhancement, because the kinetic terms should be well-defined in there, and since we are in a $w=3$ limit it cannot be an infinite distance boundary either. So we conclude that, in the M-theory frame, as $\phi \to \infty$ we reach a boundary $\CM_{{\rm 5d}}^{\rm VM}$ of the type (iii). This is indeed what happens in all the examples of $w=3$ limits analysed in section \ref{s:examples}, where any ${\cal D}_\sig$ that lies in $\ker {\bf K}$ is an effective generalised del Pezzo divisor.\footnote{It would be interesting to verify this expectation in full generality, perhaps using the recent approach in \cite{Gendler:2022ztv} to describe the cone of effective divisors in terms of GV invariants. } As such, it has non-trivial triple intersection number, implying a non-trivial scalar curvature \eqref{Rrigid} for the rigid field theory and a divergence in the scalar curvature $R_{\rm IIA}$ measured in Planck units.

\subsection{The curvature criterion}
\label{ss:CC}

Notice that the discussion of section \ref{ss:4dEFT} is based on two simple ingredients. The Distance Conjecture and the existence of a non-trivial rigid EFT below the SDC cut-off $m_*$. The EFT is non-trivial in the sense that it has a non-trivial field space with a non-vanishing curvature $R_{\rm rigid}$. If it remains so along the infinite distance trajectory, then the curvature divergence follows. This simple picture motivates the following proposal, that should apply to moduli spaces of vacua of EFTs compatible with quantum gravity:

\begin{conjecture}[Curvature Criterion] \

Along a geodesic trajectory of infinite distance, a moduli space scalar curvature that diverges asymptotically implies the presence of a field theory sector that is decoupled from gravity.

\label{conj:CC}
\end{conjecture}

In our $\CN=2$ setup, the decoupling is implemented via the different scaling of the kinetic terms of a rigid field theory subsector compared to the rest, including gravitational interactions. For the EFT string limits analysed above, whenever $\ker {\bf K}$ is non-trivial the kinetic terms of the rigid theory do not scale in units of $m_*$, leading to a divergence of the form 
\begin{equation}
    R \simeq \frac{M_{\rm P}^2}{m_*^2} R_{\rm rigid} \sim \phi^w \, .
    \label{Rasym}
\end{equation}
This lesson however applies beyond this particular kind of limits. Let us for instance consider the case in which, on top of the limit \eqref{limita}, the saxions that correspond to $\ker {\bf K}$  scale like $\phi^\delta$. As long as $0\leq \delta < 1$, the different scaling of the kinetic terms and the decoupling between the rigid field theory and the gravitational sector will occur, now with a different rate. When translated into the asymptotic behaviour of the moduli space metric, one will also find a curvature divergence of the form \eqref{Rasym}, except with $\phi^w$ replaced by $\phi^{w(1-\delta)}$ in the rhs. 

In this context, what Conjecture \ref{conj:CC} proposes is  
that the existence of a rigid field theory subsector with a non-vanishing curvature is the only mechanism to generate a curvature divergence along infinite distance limits. This proposal is very suggestive from the viewpoint of the original Conjecture 3 of \cite{Ooguri:2006in}, because it preserves the intuition that dominant gravitational effects should lead to a negative asymptotic curvature. Additionally, it may lead to an interesting interpretation of the smoothing surgery prescription proposed in \cite{Cecotti:2021cvv}, in which the Weil-Petersson metric of the vector multiplet moduli space $\CM_{\rm 4d}^{\rm VM}$ is replaced by the complete Hodge metric on the same manifold. Since this replacement should remove the curvature divergences, if our proposal is correct it should have an interpretation in terms of any sector of the compactification that leads to a rigid EFT. It is however not clear whether it should amount to integrate in all the massive states charged under the rigid field theory sector, as proposed in \cite{Cecotti:2021cvv}. 

Indeed, the massive spectrum above the SDC cut-off $m_*$ that is charged under the rigid sector is quite remarkable. On the one hand, we have the towers of states predicted by the dual description of the theory, which scale like $m_*$ along the limit, and that were accounted for in \cite{Lee:2019wij,Marchesano:2022axe}. On the other hand, there are additional infinite towers of states that also scale like $m_*$, but that are charged under the gauge sector that corresponds to $\ker {\bf K}$. Such towers were described for EFT string $w=3$ limits at the end of section \ref{ss:Mth}, and they have the particularity that they are charged under gauge couplings that are constant along the infinite distance limit, and therefore they have a finite physical charge. Because the mass of their states scales like $m_* \sim \phi^{-w/2}$, these towers have an extremality factor that asymptotes to infinity. Notice that this indicates a regime where gauge interactions dominate parametrically over gravitational ones, again in agreement with the presence of the rigid theory below the SDC scale. When going above $m_*$ for $w=3$ limits, we recover the strong coupling fixed points of the 5d supersymmetric gauge theories with $E_n$ global symmetries found in \cite{Seiberg:1996bd}, which can be engineered via collapsing del Pezzo divisors in M-theory and generalisations \cite{Morrison:1996xf,Douglas:1996xp,Intriligator:1997pq}. In this sense, the results in \cite{Ooguri:1996me,Lawrence:1997jr,Gopakumar:1998ii,Gopakumar:1998jq} should be crucial in checking the proposal of \cite{Cecotti:2021cvv} mentioned above. 

One may now consider infinite distance limits where the scalar curvature becomes positive asymptotically, although this time bounded from above. In our type IIA setup, these are for instance $w=3$ limits where $[\bm{e}] \in \CM_{\rm 5d}^{\rm VM}$ lies outside of a boundary of the type (iii), but close to it. In this case all gauge couplings are asymptotically vanishing at the same rate, but there is still a hierarchy between gauge sectors. The coupling $g_\sig$ that corresponds to a small del Pezzo divisor ${\cal D}_\sig$ will be much stronger than a coupling related to any Nef divisor, and the tower of states made of D2-branes wrapping curves of ${\cal D}_\sig$ will be much lighter than other D2-brane towers, dominating the massive spectrum below the species scale. This again implies a gauge subsector that dominates over the remaining gauge interactions and the gravitational ones, although not parametrically like when $[\bm{e}]$ lies in the boundary of $\CM_{\rm 5d}^{\rm VM}$. From the viewpoint of the 5d gauge theories of \cite{Seiberg:1996bd} the limit specifies a point in the Coulomb branch away from the fixed point. 

Interestingly, our results from sections \ref{s:limits} and \ref{s:examples} suggest that the only way to obtain an asymptotically positive scalar curvature is to consider a limit that is close to a divergent curvature limit, in agreement with the spirit of Conjecture \ref{conj:CC}.


\section{The curvature along different limits}
\label{s:limits}

In the following we analyse in more detail the dependence of the scalar curvature \eqref{scalarIIA} along the different  large-volume EFT string limits \eqref{limita}, in order to substantiate the picture presented in the last section. A first question is whether or not the saxion-dependent quantity \eqref{Y2IIAtil} tends to infinity along a given limit. This can be addressed by diagonalising the gauge kinetic matrix $\tilde{I}_{ab}$, and finding how each of its eigenvalues depend on the parameter $\phi$. While in general $\tilde{I}_{ab} = \frac{2}{3} \CK g_{ab}$ leads to moduli-dependent eigenvectors and eigenvalues, one can focus on their leading behaviour with respect to $\phi$, which turns out to be specified by topological data. One then obtains an approximate eigenvector expression
\be
u^a_M(\phi) = v^a_M + \CO(\phi^{-1})\, ,
\label{eigen}
\ee
where $\bm{v}_M$ is $\phi$-independent  and has integer entries, so it can be identified with a divisor class ${\cal D}_M$ of $X$.  When contracted with the kinetic matrix, one can extract its asymptotic behaviour along this particular direction
\be
\tilde{I}_{MM} \equiv \tilde{I}_{ab}v_M^a v_M^b \sim \phi^{w_M} \, , \qquad w_M \in \mathbb{Z} \, ,
\label{weightM}
\ee
which captures the leading asymptotic behaviour of the corresponding eigenvalue. Let us call $w_M$ the weight of the vector $v_M$.  Plugged back into \eqref{Y2IIAtil} one finds that
\be
{\cal Y }^2 \sim \sum_{M, N, L} \left( v_M^a v_N^b v_L^c \CK_{abc}\right)^2 \phi^{w - w_M - w_N - w_L} \, ,
\label{Y2sum}
\ee
from where one can determine if the curvature diverges or not. The details of this computation are described in Appendix \ref{ap:details}, and it turns out that the result essentially depends on the  vector $\bm{e}$ specifying the limit and its contractions with $\CK_{abc}$. Therefore the calculation varies for each class $w=1,2,3$ of limits, as we discuss below. 

Since $\tilde{I}_{ab}$ is a Hodge star metric, one can give a more accurate description of its asymptotic diagonalisation by using the same techniques as in \cite{Grimm:2019ixq,Gendler:2020dfp}. This method is more effective the less degenerate the limit is, in the sense of \cite{Grimm:2018cpv} and \cite{Lanza:2021udy}. As discussed in Appendix \ref{ap:degenerate}, limits with scaling weight $w=3$ are maximally degenerate, those with $w=2$ are partially degenerate and $w=1$ corresponds to non-degenerate limits. We thus apply the mixed Hodge structure approach of \cite{Grimm:2019ixq,Gendler:2020dfp} to $w=2$ and $w=1$ limits. Remarkably, we find that in a large fraction of limits the (classical) scalar curvature is asymptotically negative and has a universal behaviour.


\subsection{M-theory limits}
\label{ss:Mtheory}

We first analyse type IIA limits with scaling weight $w=3$, which we also dub as M-theory limits. As explained around \eqref{Mtrajw3}, they correspond to a decompactification limit to M-theory on $X$, and their endpoints have a particularly simple description in terms of M-theory variables. 

Following the strategy described above, we consider the diagonalisation of the kinetic matrix $\tilde{I}_{ab} = \frac{2}{3} \CK g_{ab}$ along a $w=3$ limit \eqref{limita}. For this we expand the elements that define the metric in powers of $\phi$ 
\be
\begin{split}
&\CK =  {\bf k} \phi^3 + 3{\bf k}_a t^a_0 \phi^{2} + 3{\bf K}_{ab}t^a_0 t^b_0 \phi + \CK_{abc} t^a_0 t^b_0 t^c_0\, ,\\
&\CK_a =  {\bf k}_a \phi^2 + 2{\bf K}_{ab} t^b_0 \phi + \CK_{abc} t^b_0 t^c_0 \, ,\\
&\CK_{ab} =  {\bf K}_{ab}  \phi + \CK_{abc} t^c_0\, ,
\end{split}
\label{Kexpand}
\ee
%
where ${\bf k} \equiv \CK_{abc} e^ae^be^c$, ${\bf k}_a \equiv \CK_{abc} e^be^c$, and ${\bf K}_{ab} = \CK_{abc}e^c$ have been defined as in section \ref{s:typeIIA}. From the results of Appendix \ref{ap:details} one can see that there are three classes of leading vectors in the expansion \eqref{eigen}, which can be defined as
\be     \label{basisw3}
\bm{v}_0 \in \ker {\bf K}  \, , \qquad v_I^a {\bf k}_a = 0 \text{ but } \bm{v}_I \notin \ker {\bf K}\, , \qquad  \bm{v}_{e}  = \bm{e}  \, ,
\ee
and that the weight \eqref{weightM} for each kind of eigenvector is
\be
w_0 = 0, \qquad w_I = 1, \qquad w_{e} = 1 .
\label{ww3}
\ee
Notice that a vector of the form $\bm{v}_{I}$ can be shifted by one in the class $\bm{v}_0$ without spoiling its definition, but as long as  $w_{0} < w_I$ the weights are well-defined. Also, the weight $w_e = w-2$ follows from the general properties of EFT string, that constrain its tension ${\cal T}_e = 3 e^a \CK_a/2\CK$ to asymptote like $\phi^{-1}$ along its own EFT string flow \cite{Lanza:2021udy}. It follows from \eqref{ww3} that we can only have a divergent term in \eqref{Y2sum} whenever a vector of the form $\bm{v}_{0}$ is involved, which implies that the matrix ${\bf K}$ defined in \eqref{rank} must be singular. In the language of \cite{Corvilain:2018lgw}, this means that along limits of the form IV$_{n_V}$ the curvature does not diverge, but along those of the type IV$_{d<n_V}$  it will, provided that the elements within $\ker {\bf K}$ have a non-vanishing triple intersection. This is the case for generalised del Pezzo divisors, which at the end of section \ref{ss:Mth} were identified with the condition $\bm{v}_{0} \in \ker {\bf K}$. Such divisors define a rigid field theory with classical prepotential 
\be
\cF_{\rm rigid}^{\rm cl} = -\frac{1}{6} \cK_{\a\b\g} T^\a T^\b T^\g  -\frac{1}{6} \cK_{\a\b\Gamma} T^\a T^\b T_0^\Gamma -\frac{1}{6} \cK_{\a\Sigma\Gamma} T^\a T_0^\Sigma T_0^\Gamma -\frac{1}{6} \cK_{\Lambda\Sigma\Gamma} T_0^\Lambda T_0^\Sigma T_0^\Gamma   \, , 
\label{prepw3}
\ee
where ${\cal D}_\a$ form an integer basis of $\ker {\bf K}$, and   ${\cal D}_\Gamma$ form a basis for the vectors of the form $\bm{v}_I$. Some examples of this rigid prepotential are worked out in section \ref{s:examples} and in  \cite[section 4]{Alexandrov:2017mgi}. 

One can also see in a more direct manner that the curvature does not diverge when $\ker {\bf K}$ is trivial. Indeed, let us consider the explicit expression for the inverse of $\tilde{I}_{ab}$
\be
\tilde{I}^{ab} = 3 \frac{t^at^b}{\cK} -  \cK^{ab}\, ,
\label{invI}
\ee
where $\CK^{ab}$ is the inverse of $\CK_{ab}$. Plugging this expression into \eqref{Y2IIAtil} we obtain
\be
{\cal Y }^2 =  \frac{3}{2} n_V -  \CV_X  \CK^{ab}\CK^{cd}\CK^{ef}  \CK_{ace} \CK_{bdf} \, .
\label{Y2alt}
\ee
If ${\bf K}_{ab}$ is invertible with inverse ${\bf K}^{ab}$, then we have that
\be
\CK^{ab} = {\bf K}^{ab} \phi^{-1} + \CO(\phi^{-2})\, ,
\ee
and so
\be
{\cal Y }^2 =  \frac{3}{2} n_V -  \CV_X  \phi^{-3} {\bf K}^{ab} {\bf K}^{cd} {\bf K}^{ef}  \CK_{ace} \CK_{bdf} + \CO(\phi^{w-4}) \, .
\label{Y2alt2}
\ee
The second term scales like $\phi^{w-3}$ and so it will only asymptote to a non-vanishing constant for $w=3$ limits. Because ${\bf K}$ is the intersection matrix of curves within the divisor ${\cal D}_{\bm{e}}$, it has signature $(1, r-1)$, so most of its eigenvalues are negative while its trace is a positive integer. Therefore, whenever ${\bf K}$ has a very small eigenvalue $\lambda_\sig$ it must be negative, and one can estimate the second term in \eqref{Y2alt2} by $\frac{\bf k}{6} |\lambda_\sig|^{-3}$. One expects that ${\cal Y}^2$ is asymptotically finite but large when the vector $\bm{e}$ is close to a vector for which ${\bf K}$ is singular. To make this intuition more precise let us rewrite \eqref{Y2alt2} as
\be
{\cal Y }^2 =  \frac{3}{2} n_V -   {\bf K}_M^{ab} {\bf K}_M^{cd} {\bf K}_M^{ef}  \CK_{ace} \CK_{bdf} + \CO(\phi^{w-4}) \, ,
\label{Y2alt3}
\ee
where ${\bf K}_M^{ab}$ is the inverse of ${\bf K}^M_{ab} \equiv  (6/{\bf k})^{1/3} \CK_{abc}e^c$. The eigenvalues of ${\bf K}^M_{ab}$ depend continuously on the M-theory parametrisation \eqref{Mtrajw3}, and so ${\cal Y }^2$ will be large when $[\bm{e}] \in \CM_{\rm 5d}^{\rm VM}$ is close in the M-theory moduli space metric to a singularity in ${\bf K}_M^{ab}$. This locus coincides with a singularity of ${\cal I}_{ab}$, or in other words with a boundary of type (iii), as expected from our previous discussion.


\subsection{F-theory limits}
\label{ss:Ftheory}

Let us consider $w=2$ limits, that correspond to decompactification limits to F-theory on $X$, focusing first on the case of a smooth elliptic fibration. Following the convention in \cite[section 3.4]{Corvilain:2018lgw} and \cite[section 4]{Marchesano:2022axe}, we take a K\"ahler cone basis $\omega_a =\{\omega_E, \omega_\a\}$  where $\omega_\a = \pi^* \omega'_\a$ is the pull-back of a simplicial K\"ahler cone basis of $B_2$. Their triple intersection numbers are
\begin{equation}
    \CK_{EEE} = \eta_{\a\b}c_1^\a c_1^\b, \qquad \CK_{EE\a} = \eta_{\a\b} c_1^\b, \qquad \CK_{E\a\b} = \eta_{\a\b}, \qquad \CK_{\a\b\g} = 0,
    \label{interfib}
\end{equation}
where $c_1(B_2) = c_1^\a \omega'_\a$ and $\eta_{\a\b}$ is a symmetric matrix with signature $(1, h^{1,1}(B_2) -1)$. We now obtain a $w=2$ limit by taking a vector $\bm{e}$ of the form $(0, e^\a)$, $e^\a \in \mathbb{N}$. Then, one can again consider the expansion \eqref{Kexpand}, but now since ${\bf k} =0$ and 
\begin{equation}    \label{basisw2}
    {\bf k}_a =  \eta_{\a\b} e^\a e^\b \, \delta_{aE}, \qquad {\bf K} =
    \begin{pmatrix}
   \eta_{\a\b} e^\a c_1^\b & \eta_{\a\b} e^{\b} \\  \eta_{\a\b} e^{\b} & 0
    \end{pmatrix} ,
\end{equation}
the classes of leading eigenvectors in \eqref{eigen} read
\be
\bm{v}_0 \in \ker {\bf K} = \{ (0, f^\b)\, |\, \eta_{\a\b} e^\a f^\b =0\}  \, , \qquad \bm{v}_e = \bm{e} =  (0, e^\a) \, , \qquad  \bm{v}_{E} =  (1, \vec{0}) \, ,
\label{eigenw2}
\ee
see Appendix \ref{ap:details}. Working out the metric expansion one also deduces that 
\be
w_0 = 0\, , \qquad w_e = 0\, , \qquad w_E = 2\,  ,
\label{ww2}
\ee
so naively one could have divergent terms in the sum \eqref{Y2sum}, if only vectors of the form $\bm{v}_0$ and $\bm{v}_{e}$ are involved. However, because all non-vanishing triple intersection numbers involve at least one vector in $\bm{v}_{E}$, one cannot build a divergent term, and so ${\cal Y}^2$ must be asymptotically constant.

It turns out that for smooth elliptic fibrations one cannot only show that ${\cal Y}^2$  does not diverge along $w=2$ limits, but also that \eqref{scalarIIA} asymptotes to a specific negative value. To see this, let us apply the approach used in \cite{Grimm:2019ixq,Gendler:2020dfp} to the present setup. Given the triple intersection numbers \eqref{interfib}, we have that the K\"ahler potential $K = -\log (\frac{4}{3}\CK)$ is determined by
\be
\CK  = \eta_{\a\b}c_1^\a c_1^\b t_E^3 + 3 \eta_{\a\b}c_1^\a t^\b t_E^2 + 3 \eta_{\a\b} t^\a t^\b t^E .
\ee
Following the philosophy of \cite{Grimm:2019ixq,Gendler:2020dfp}, we consider the following limit
\be
t^\a = t^\a_0 + \phi^\del + e^\a \phi, \qquad t^E = t^E_0  , \qquad \del \in (0,1) .
\label{w2limit}
\ee
This represents a growth sector in which the K\"ahler direction along $\bm{e}$ grows much faster than any other direction of the base, which in turn grow faster than the K\"ahler coordinate $t^E$ that measures the area of the elliptic fibre and that it pairs up with $\bm{v}_E$ in the K\"ahler form. The mixed Hodge structure approach used in \cite{Grimm:2019ixq,Gendler:2020dfp} allow us to keep an approximate form of the K\"ahler potential to describe the metric along this growth sector, namely
\be
K = - \log \left(4 \eta_{\a\b} t^\a t^\b t^E + \dots  \right) \, , 
\label{Kgs1}
\ee
where we have neglected subleading terms in $\phi$. The metric that one derives from this approximate expression reads 
\be
g = \begin{pmatrix}
 \frac{1}{4(t^E)^2} \\ & \frac{\eta_\a\eta_\b}{\eta^2} - \frac{\eta_{\a\b}}{2\eta}
\end{pmatrix}\, ,
\ee
where $\eta_\a = \eta_{\a\b}t^\b$ and $\eta = \eta_{\a\b}t^\a t^\b$. We therefore find a gauge kinetic matrix of the form
\be
\tilde{I} = 
\begin{pmatrix}
 \frac{\eta}{2t^E} \\ & t^E \left(2\frac{\eta_\a\eta_\b}{\eta} - \eta_{\a\b}\right) 
\end{pmatrix},
\label{tIw2}
\ee
up to subleading terms. Remarkably, the result is independent of the value of $\delta$ and, because $\eta_{\a\b}$ is non-degenerate, reproduces the weights \eqref{ww2}. The inverse matrix at leading order reads
\be
\tilde{I}^{-1} = 
\begin{pmatrix}
 \frac{2t^E}{\eta} \\ & \frac{1}{t^E} \left(2\frac{t^\a t^\b}{\eta} - \eta^{\a\b}\right) 
\end{pmatrix} ,
\label{tIinvw2}
\ee
matching the results of Appendix \ref{ap:details}. From here we obtain
\be
{\cal Y}^2 = 3(n_V-1) + \CO (\phi^{-1})\, .
\label{Y2w2}
\ee
As a result, we find a scalar curvature with the negative asymptotic value 
\be
R_{\rm IIA}^{\rm cl}/2\ = - n_V^2 + 2n_V - 3 +\CO (\phi^{-1})\, .
\label{Rneg}
\ee

Now, recall that \eqref{Rneg} is the curvature associated to the classical prepotential $\cF^{\rm cl}$. The fact that it is negative  corresponds to a rigid theory with no curvature at the classical level.  Indeed, in the case at hand, the rigid $\CN=2$ theory has as dynamical fields the K\"ahler variables $\tilde{T}^\b$ that correspond to the vectors in $\bm{v}_{0}$. By construction, these are divisors which are pullbacks from curves of the base with negative definite self-intersection matrix $\tilde{\eta}_{\a\b}$. The classical prepotential of the rigid theory reads
\be
\cF_{\rm rigid}^{\rm cl} =  \frac{1}{2} \tilde{\eta}_{\a\b} \tilde{T}^\a \tilde{T}^\b T_0^E + \dots
\label{prepw2cl}
\ee
where the dots contain at most linear terms on the dynamical fields. Since there is no cubic term, from \eqref{Rrigid} we have that $R_{\rm rigid}^{\rm cl} = 0$, in agreement with the fact that \eqref{prepw2cl} corresponds to a flat classical metric $g_{\a\b}^{\rm cl} = T_0^E\tilde{\eta}_{\a\b}$. Polynomial curvature corrections in \eqref{fullF} will not change this result, but exponential world-sheet corrections could do so. Indeed, whenever $\ker {\bf K}$ is non-trivial we have a set of vertical divisors ${\cal D}_\beta$ that do not intersect ${\cal D}_e$. Projected to the base $B_2$, these lead to a set of curves ${\cal C}_\b= \pi({\cal D}_\beta)$ that do not intersect $\pi({\cal D}_e)$ in the base. As such they must have negative self-intersection and a finite number of nonzero genus zero GV invariants that will contribute to \eqref{fullF}. Unless all GV invariants vanish, this will lead to terms that are independent of $\phi$ if we take $\delta=0$ in \eqref{w2limit}. Even if exponentially suppressed for large values of $t^\b_0$, they will give a non-vanishing, $\phi$-independent contribution to $\p_\b \p_\g \p_\delta \cF$, where all indices belong to $\ker {\bf K}$. Therefore, whenever $\ker {\bf K}$ is non-trivial and its projection to the base leads to curves ${\cal C}_\b$ with non-vanishing GV invariants, we expect to have a curvature divergence along $w=2$ EFT string limits \eqref{limita}, sourced by quantum corrections in the rigid field theory.\footnote{It may seem surprising that world-sheet instanton corrections can modify the large-volume asymptotic curvature, since under the assumption $t_0^a \gg 1$ one can neglect them to compute the metric. Note however that the curvature also depends on derivatives of the metric, and that \eqref{metric} has plenty of axionic isometries. Only world-sheet instanton terms break such isometries, so they are indeed the ones that are able to correct the curvature.} It would be important to test this expectation explicitly and, if true, to understand it from an F-theory viewpoint.

Let us now consider non-smooth elliptic fibrations, whose triple intersection numbers do not correspond to \eqref{interfib}. In the case that $r = $ rank ${\bf K} = n_V$, we can use \eqref{Y2alt2} to deduce that
\be
R_{\rm IIA}^{\rm cl}/2\ = - n_V^2 + \oh n_V + \CO (\phi^{-1})\, .
\label{Rmax}
\ee
The remaining cases are $w=2$ limits which, in the nomenclature used in \cite{Corvilain:2018lgw}, correspond to type III$_c$ limits with $0<c< n_V-2$, and so ${\bf K}$ has a rank larger than 2. The classes of leading eigenvectors \eqref{eigen} that  generalise \eqref{basisw2} are
\be     \label{basisw20}
\bm{v}_{0} \in \ker {\bf K}  \, , \qquad  \bm{v}_e =  \bm{e}  \, , \qquad  {v}_{I}^a {\bf k}_a = 0 \ \text{but}\  \bm{v}_I \notin \ker {\bf K} \oplus \langle \bm{e} \rangle \, , \qquad {v}_{E}^a {\bf k}_a \neq 0 \, .
\ee
As in \eqref{basisw3}, the split between these classes of vectors is not unique, but the weights are well defined as long as $w_0, w_e < w_I < w_E$, which should be the case. More precisely we expect 
\be
w_0 = 0\, , \qquad w_e = 0\, , \qquad w_{I} = 1\, ,  \qquad w_{E} = 2\, .
\label{ww20}
\ee
To see this notice that the vectors $\bm{v}_0$, $\bm{v}_e$ represent vertical divisors as in the smooth fibration case, and $w_0 = w_e = 0$ follows from general arguments. The vector $\bm{v}_E$ again  represents the dual divisor to the generic fibre, and so again $w_E =2$ is expected. Finally, the vectors of the form $\bm{v}_I$ represent the divisors that arise from fibral degenerations, and their weight is motivated by examples that we analyse in section \ref{s:examples}, although it would be nice to confirm its value in general. 

With this set of weights one cannot have asymptotic divergences for ${\cal Y}^2$, even for this more general class of $w=2$ limits. Indeed, the intersection of two vertical divisors is always proportional to the generic fibre, and as a result in the sum \eqref{Y2sum} two vectors of the form $\bm{v}_{0}$ or $\bm{v}_e$ should always combine with a vector $\bm{v}_E$, as in the smooth case. Additionally, we can have one vector $\bm{v}_{0}$ or $\bm{v}_e$ coupling to two vectors $\bm{v}_I$, but this combination also kills the potential divergence.  Therefore, we expect an asymptotic behaviour for ${\cal Y}^2$ that is analogous to the smooth fibration and maximal rank cases. A natural guess for this value is
\be
{\cal Y}^2 = 3(n_V -  r/2) + \CO(\phi^{-1})\, .
\ee
It would be interesting to test this guess with explicit examples. Again, whenever $r < n_V$ it could be that world-sheet instanton corrections generate a non trivial $R_{\rm rigid}$, and therefore a divergence for the asymptotic scalar curvature, if the appropriate GV invariants are non-trivial.

Finally, let us consider the tower of states that are charged under the gauge group of the rigid theory. In $w=2$ EFT string limits we have two universal towers of asymptotically massless states that combine into a lattice. These are D0-branes, D2-branes wrapping the generic fibre, and bound states of them. The gauge kinetic matrix of this lattice blows up along the limit, signalling a decompactification limit to F-theory. However, as pointed out in \cite[section 4]{Marchesano:2022axe} there can be a larger lattice of light states that tend to zero mass as fast as the D0-brane tower. To see this, let us consider an effective vertical divisor ${\cal D}_\sig$ of $X$ that corresponds to a vector of the form $\bm{v}_{0}$ in \eqref{eigenw2}. If we set $\delta=0$ in \eqref{w2limit}, the volume of this divisor will remain constant as $\phi \to \infty$, and so the DBI action of a D4-brane wrapped around such divisor only depends on $\phi$ through the 10d dilaton dependence, just like D0-branes. The same is true if we add an Abelian worldvolume flux ${\cal F}$ threading the internal D4-brane worldvolume. Scanning over the worldvolume flux quanta, we generate an infinite lattice of D4/D2/D0 BPS bound states. Such a lattice is charged under $U(1)$'s with constant gauge couplings, and so we again have a tower of particles with an extremality factor $\g$ that asymptotes to infinity. This infinite lattice disappears once that we analyse the system from the viewpoint of F-theory compactified on $X$, in contrast with the towers with $\g \to \infty$ of $w=3$ limits in the M-theory picture.


\subsection{Emergent string limits}
\label{ss:critical}

Let us finally consider $w=1$ limits, that should correspond to emergent string limits. In this case we have that ${\bf k} = {\bf k}_a =0$ and
\begin{equation}
{\bf K} =
    \begin{pmatrix}
   0 & 0 \\ 0 & Q_{pq}
    \end{pmatrix} ,
\end{equation}
where the first entry corresponds to $\bm{e}$. Geometrically, $\bm{e}$ represents a Nef divisor class ${\cal D}_e$ whose topology is either that of a $K3$ or a $T^4$, and $Q$ determines the lattice dual to $\Lambda^\vee = \iota^* H^2(X_6, \mathbbm{Z})$, where $\iota$ is the embedding map of such surface into $X_6$ \cite{Lee:2019wij}. Using that all non-proportional Nef divisors should intersect, one sees that the quadratic form ${\bf Q}$ is such that all K\"ahler moduli besides  $t^a \propto e^a$ appear in $Q_{pq}t^p t^q$, although this does not mean that $r = {\rm rank}\, {\bf Q}$ has the maximal value $n_V-1$.  In the classification of \cite{Corvilain:2018lgw} these are limits of type II$_r$. 
 
To describe the asymptotic behaviour of the metric it is useful to consider the same classes of eigenvectors as in \cite[eq.(5.14)]{Marchesano:2022axe}:
\be
 \bm{v}_{e} = \bm{e} \, , \qquad  \bm{v}_{0} \in \ker {\bf K}\, , \qquad \bm{v}_{I} \notin \ker {\bf K}\, , 
 \label{eigenw1}
\ee
where the $\bm{v}_{0}$'s are linearly independent from $\bm{e}$. Then from expanding the gauge kinetic matrix in powers of $\phi$ one obtains that 
\be
w_{e} = -1, \qquad w_0 = 0, \qquad w_{I} = 1 ,
\label{ww1}
\ee
in agreement with the results of \cite{Marchesano:2022axe}. Again, the value of $w_{e}$ is fixed by EFT string asymptotic behaviour ${\cal T}_e \sim \phi^{-1}$ along its own limit. While this negative power could be a naive source of divergences for ${\cal Y}^2$, the structure of triple intersection numbers implies that each non-vanishing element of the sum \eqref{Y2sum} that contains $\bm{v}_e$ must also contain two vectors $\bm{v}_I$, and so such a combination cannot be a source of divergence. One may worry that quantum corrections can modify this result, but quantum-corrected cubic terms that involve the vector $\bm{e}$ have an exponential suppression of the form $\exp(-2\pi k\phi)$, and so they become irrelevant at $\phi \to \infty$. A linear divergence for the scalar curvature only arises if we have vectors or the form $\bm{v}_{0}$ with non-trivial triple intersection numbers, which requires to consider limits with $r < n_V-1$. 

If $\ker {\bf K} = \langle \bm{e}\rangle$ (i.e., $r=n_V-1$) one can get a precise expression for the asymptotic value of the scalar curvature. As shown in \cite{Lanza:2021udy} and Appendix \ref{ap:degenerate}, $w=1$ limits are non-degenerate, which makes them particularly suitable for applying the mixed Hodge structure techniques that we applied to $w=2$ limits. In this case the limit \eqref{limita} corresponds to the growth sector $t \gg \tilde{t}^p$, where $t$ corresponds to the K\"ahler modulus along $\bm{e}$ and $\tilde{t}^p$ to the remaining ones, among which no growth hierarchy is needed to pursue the analysis. The approximate K\"ahler potential reads
\be
K = - \log \left(4 t Q_{pq} \tilde{t}^p \tilde{t}^q + \dots  \right) \, , 
\label{Kw1}
\ee
where the terms that we are neglecting are $\phi$-independent. Due to the similarity with \eqref{Kgs1} everything works as in the smooth-fibration $w=2$ limit, except that now it could be that ${\bf Q}$ has a non-trivial kernel. We find that
\be
\tilde{I} = 
\begin{pmatrix}
 \frac{Q}{2t} \\ & t\left(2\frac{Q_p Q_q}{Q} - Q_{pq}\right)
\end{pmatrix}
\sim
\begin{pmatrix}
 \phi^{-1} \\ & \phi \\ & & \ddots
\end{pmatrix} ,
\ee
where $Q_p = Q_{pq}\tilde{t}^q$ and $Q = Q_{pq}\tilde{t}^p \tilde{t}^q$. The dots represent saxion directions in the kernel of ${\bf Q}$, which do not scale with $\phi$. We therefore reproduce the results in \eqref{ww1}, and find again that the elements of $\ker {\bf Q}$ are the only possible sources of divergence for ${\cal Y}^2$. Notice moreover that when $\ker {\bf Q}$ is trivial the leading kinetic matrix $\tilde{I}$ is completely analogous to that in \eqref{tIw2}, in the sense that its inverse looks like \eqref{tIinvw2}, with the replacement $\bm{\eta} \to {\bf Q}$. The computation of ${\cal Y}^2$ works exactly the same as in that case, and at the end one obtains an asymptotically constant and negative curvature of the form \eqref{Rneg} for all these type of limits. 


\subsection{Summary}
\label{ss:summary}

Let us summarise the results of this section, which are gathered in table  \ref{tab:curv(w,rk)}. The general pattern is that the scalar curvature does not diverge along the EFT string limits \eqref{limita} when the rank of the matrix ${\bf K}_{ab} = \CK_{abc}e^c$ is maximal, while it may diverge as $\CV_X \sim \phi^{w}$ for  the other cases. The physical source of divergence is the curvature $R_{\rm rigid}$ of the rigid field theory below the SDC scale $m_* \simeq m_{\rm D0}$. Such a curvature vanishes at the classical level for $w=2$ limits, and that is why $R_{\rm IIA}^{\rm cl}$ does not diverge asymptotically even when $r < n_V$. The presence of non-trivial world-sheet instantons in the rigid sector is expected to correct this result to:
\begin{equation}
    R_{\rm IIA}(w=2, r< n_V) \sim  \cO(e^{-t_0^\sig}) {\bf k}_\Sigma t^\Sigma_0 \phi^2\, ,
    \label{quantumRw2}
\end{equation}
where $t^\sigma_0$ are dynamical saxion and $t^\Sigma_0$ frozen saxion vevs from the viewpoint of the rigid theory. This behaviour will only occur if there is at least a curve ${\cal C}$ of the elliptic fibration base whose area remains constant along the limit and it has a non-vanishing GV invariant. Besides this, the effect of world-sheet instantons is to correct $R^{\rm cl}_{\rm rigid}$ by a term of order $\cO(e^{-t_0^\Sigma})$.

\setlength{\arrayrulewidth}{0.2mm}
\renewcommand{\arraystretch}{1.5}
\begin{table}[H]
\begin{center}
\begin{tabular}{|cl|c|}
\hline
\multicolumn{2}{|c|}{Case} & $R_{\rm IIA}^{\rm cl}(\phi \to \infty)$  \\
\hline\hline
\multirow{2}{*}{$w$=3} & $r=n_V$ & $- 2n_V^2 + n_V + \mathfrak{C}$ \\
\cline{2-3}
& $r < n_V$ & $ R_{\rm rigid}^{\rm cl} \frac{2}{3}{\bf k} \phi^3$  \\
\hline
\multirow{3}{*}{$w$=2} &$ r =n_V$ & $-2n_V^2 +n_V$   \\
\cline{2-3}
& $r<n_V$, smooth & $-2(n_V^2 - 2n_V +3)$   \\
\cline{2-3}
& $r<n_V$, non-smooth &  $\left[ -2n_V^2 + 4n_V -3r\right]^*$  \\
\hline
\multirow{2}{*}{$w$=1} & $r=n_V-1$ & $-2(n_V^2 - 2n_V +3)$ \\
\cline{2-3}
& $r<n_V-1$ & $R_{\rm rigid}^{\rm cl} 2{\bf K}_{\Sigma \Lambda}t^\Sigma_0t^\Lambda_0 \phi$   \\
\hline
\end{tabular}
\end{center}
\caption{Asymptotic behaviour of the moduli space scalar curvature along  type IIA vector multiplet limits \eqref{limita}. $R_{\rm IIA}^{\rm cl}$ corresponds to \eqref{scalarIIA}, the curvature of the large-volume  metric \eqref{metric} and $R^{\rm cl}_{\rm rigid}$ to \eqref{Rrigid}. The remaining quantities are  topological data defined as $n_V = h^{1,1}(X)$, ${\bf k} \equiv \CK_{abc} e^ae^be^c$, ${\bf k}_a \equiv \CK_{abc} e^be^c$, ${\bf K}_{ab} = \CK_{abc}e^c$ and $r = {\rm rank }\, {\bf K}$, and  frozen saxions in the rigid theory $t^\Sigma_0$. The entry marked as $[ \dots ]^*$ is the result of guesswork.   \label{tab:curv(w,rk)}}
\end{table}

If the scalar curvature does not diverge along the limit it is asymptotically negative, with the exception of $w=3$ limits with $r=n_V$, where the moduli-independent quantity  
\be
\mathfrak{C} = -\frac{\bf k}{6} {\bf K}^{ab} {\bf K}^{cd} {\bf K}^{ef}  \CK_{ace} \CK_{bdf}
\label{frakC}
\ee
can be arbitrarily large and positive. As explained in section \ref{ss:Mth} this happens precisely when the vector $\bm{e}$, interpreted as a point in the M-theory moduli space $\CM_{\rm 5d}^{\rm VM}$, is close to a boundary where the gauge kinetic function $\cI_{ab}$ degenerates and the curvature $R_{\rm IIA}$ diverges. In practice, this large value for \eqref{frakC} is attained when ${\bf K}$ has a hierarchy of eigenvalues. Physically, this corresponds to a hierarchy between gauge couplings in the vector multiplet sector, and in particular some couplings that are much larger than the one of the graviphoton. The quotient of couplings does however not grow as $\phi \to \infty$. That would correspond to the case $r < n_V$, where we recover a rigid $\CN=2$ theory below $m_*$ for any value of $\phi$, and the curvature blows up. 

Let us stress that the content of table \ref{tab:curv(w,rk)} is only valid for limits of the form \eqref{limita} under the assumption $t_0^a \gg 1$. For other infinite distance limits there may be other ways different from a non-trivial $\ker {\bf K}$ in which one can decouple a subset of the moduli from the trajectory of infinite distance, and one should define the rigid field theory accordingly. 
A clear example are $w=1$ limits that lead to Seiberg-Witten points, which have been recently studied in \cite{Lee:2019wij,Bastian:2021hpc}. As pointed out in \cite{Bastian:2021hpc}, in the region in which the limit is defined one typically has to add some world-sheet instanton corrections in order for the metric not to be degenerate, and this automatically implements a hierarchy between different moduli. It would be very interesting to check if Conjecture \ref{conj:CC} is valid along these and other limits of the vector multiplet moduli space. 

Finally, note that a positive scalar curvature does not tell us anything about the sign of the sectional curvatures. Indeed, it was shown in \cite{Lanza:2021udy} that along EFT string limits at least one holomorphic sectional curvature must be negative asymptotically. In principle one can study the asymptotic behaviour of sectional curvatures and the Ricci tensor by considering \eqref{Riemann} and \eqref{Ricci} and applying similar techniques to the ones used in this section. It would be interesting to see which pattern of asymptotic sectional curvatures allows to have a divergent scalar curvature and at the same time a moduli space of finite volume, as it is the case for the type IIA vector moduli space with Weil-Petersson metric \cite{Todorov:2004jg,Lu:2005bi,Lu:2005bj}. We leave these questions for future work. 


\section{Examples}
\label{s:examples}

In this section we work out some examples that illustrate our discussion and results of the previous sections. On each example we describe the asymptotic behaviour of the type IIA scalar curvature $R_{\rm IIA}$ in terms of the associated M-theory vector moduli space. For those limits where the curvature blows up, we reproduce the divergence from the curvature of the associated rigid theory. Our examples correspond to Calabi--Yau manifolds with two or three moduli, but in some cases the K\"ahler cone needs to be extended to construct the full set of limits. As discussed in section \ref{s:curvature}, curvature divergences can appear for vectors $\bm{e}$ that lie in the boundary of the (extended) K\"ahler cone. In these examples with few moduli, they oftentimes corresponds to generators of the cone, and more precisely to elementary limits, in the sense of footnote \ref{ft:elementary}. 

\subsection*{Two-moduli example}

Let us first consider a two-modulus example based on the Calabi--Yau manifold $X = \mathbb{P}^{(1,1,1,6,9)}[18]$, which was studied in \cite{Candelas:1994hw} and can either be regarded as a weighted projective space or as an smooth elliptic fibration over $\mathbb{P}^2$. This manifold has two K\"ahler moduli and a simplicial K\"ahler cone spanned by two Nef divisors ${\cal D}_1$ and ${\cal D}_2$. In this basis the K\"ahler form reads
\begin{equation}
    J = t^1 \om_1 + t^2 \om_2\, ,
\end{equation}
where $[\om_i]$ is the Poincar\'e dual class to ${\cal D}_i$ and $t^i \geq 0$. The triple intersection numbers are
\be
\cK_{111}=9\, , \qquad \cK_{112}=3\, , \qquad \cK_{122}=1\, , \qquad \cK_{222}=0\, .
\ee
The different discrete cones relevant to infinite distance large-volume limits have been discussed in \cite[section 4.2]{Lanza:2020qmt}. In particular the cone of effective divisors is given by 
\be
\mathcal{E}=\{ (e^1,e^2)\in \mathbb{Z}^2 | e^1 \geq 0, 3e^1+e^2 \geq 0 \}\, ,
\ee
generated by the classes ${\cal D}_2$ and ${\cal D}_{B} = {\cal D}_1 - 3{\cal D}_2$, where the latter corresponds to the ${\mathbb{P}^2}$ base.

\begin{figure}[H]
\begin{center}
\includegraphics[scale=0.5]{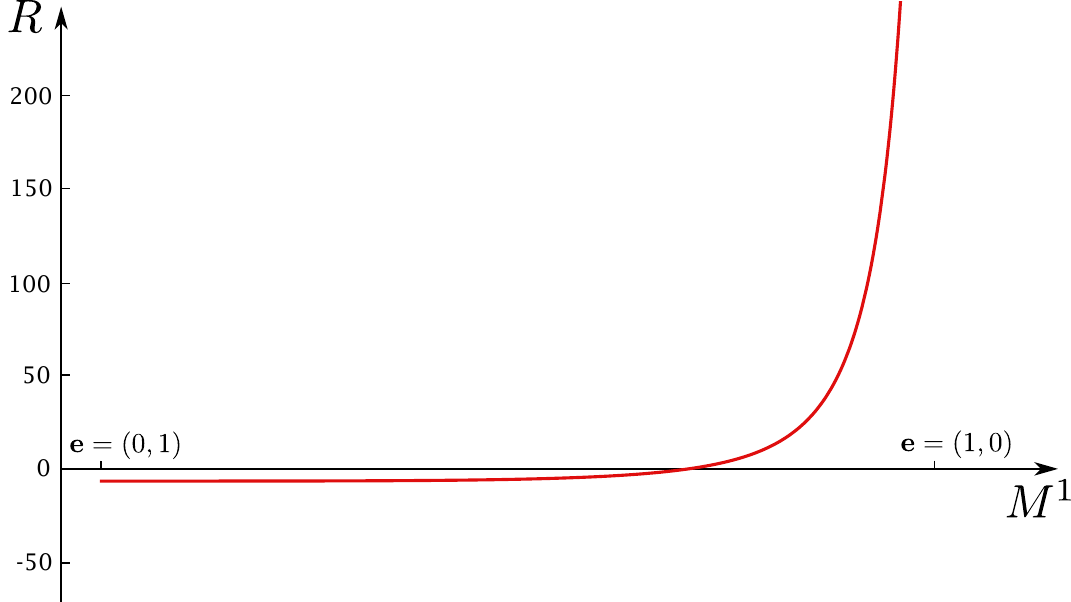}
\caption{Asymptotic behaviour of the classical scalar curvature $R^{\rm cl}_{\rm IIA}$ as a function of the M-theory modulus $M^1$ in example $X = \mathbb{P}^{(1,1,1,6,9)}[18]$.}
\label{fig:R_ex1}
\end{center}
\end{figure}

The asymptotic classical scalar curvature $R^{\rm cl}_{\rm IIA}$ along the M-theory moduli space
\begin{align}
    \cM_{{\rm 5d}, X}^{\rm VM} &  = \left\{ (M^1,M^2) \ | \ M^i \geq 0 \quad  \text{and} \quad 9(M^1)^3 +9(M^1)^2M^2 + 3 M^1(M^2)^2 =6 \right\} 
\end{align}
is depicted in figure \ref{fig:R_ex1}. One observes a divergence at $(M^1, M^2) =(\left(\frac{2}{3}\right)^{1/3}, 0)$, which corresponds to the elementary limit $\bm{e} = (1,0)$ and an asymptotic negative value at $(M^1, M^2) = (0, \infty)$, which corresponds to the elementary limit $\bm{e} = (0,1)$. Let us analyse both of these limits separately. 

\begin{itemize}

\item Limit $t^1 \to \infty$ or $\bm{e}=(1,0)$: this is a $w=3$ limit with ${\bf k} = 9$. The matrix 
\begin{equation}
    {\bf K} = 
    \begin{pmatrix}
        9 & 3 \\ 3& 1
    \end{pmatrix}
\end{equation}
has rank one and a kernel generated by $\bm{v}_{0}=(1,-3)$, that  corresponds to the effective divisor class of the base ${\cal D}_B$. In the basis $\{{\cal D}_{e},-{\cal D}_B\}$ the triple intersection numbers read\footnote{The sign flip in the effective non-Nef divisor corresponds to the fact that, from the viewpoint of the local Calabi--Yau geometry, $-{\cal D}_B$ becomes Nef. See \cite{Katz:2020ewz} for a discussion in the context of M-theory CY compactifications.} 
\be
\cK_{eee}=9 \, , \qquad \cK_{BBB}= -9\, ,
\ee
and the K\"ahler form is $J=t^e \omega_e+t^{B} \omega_{B}$, where the rotated moduli are $(t^e,t^{B})=(t^1+\frac{1}{3}t^2,\frac{1}{3}t^2)$. One can also compute the rigid prepotential \eqref{prepw3}
\begin{eqn}
    \cF_{\rm rigid}^{\rm cl} = \frac{3}{2}\left(T^{B}\right)^3\, ,
\end{eqn}
and see that $T^{B}=\frac{1}{3}T^2$ is the only field in the rigid field theory limit. The metric of this theory reads $g_{B\bar{B}} = 9t^B = 3t^2$, which leads to the curvature
\be
R_{\rm rigid} =  \frac{3}{2(t^2)^3}\, .
\label{Rrex1}
\ee
Using that $\CV_X \sim \frac{3}{2}\phi^3$ along this limit it is easy to see that \eqref{divR} reproduces the asymptotic value of the 4d supergravity moduli space classical scalar curvature
\be
R_{\rm IIA}^{\rm cl} = \frac{9}{\left(t_0^2\right)^3}\phi^3 + \mathcal{O}(\phi^2)\, .
\ee

\item Limit $t^2 \to \infty$ or $\bm{e}=(0,1)$: this is a $w=2$ limit where the matrix
\be
{\bf K} = 
\begin{pmatrix}
3 & 1 \\
1 & 0
\end{pmatrix},
\ee
has maximal rank. As such, there is no rigid theory in the IR and no scalar curvature divergence. The scalar curvature of the 4d supergravity moduli space asymptotes to a negative constant, and more precisely to
\be
R_{\rm IIA}^{\rm cl} = -6 + \mathcal{O}(\phi^{-2})\, ,
\ee
in agreement with the results of section \ref{s:limits}.

\end{itemize}

\subsection*{Two moduli with a flop}

\begin{figure}[H]
\begin{center}
\includegraphics[scale=0.5]{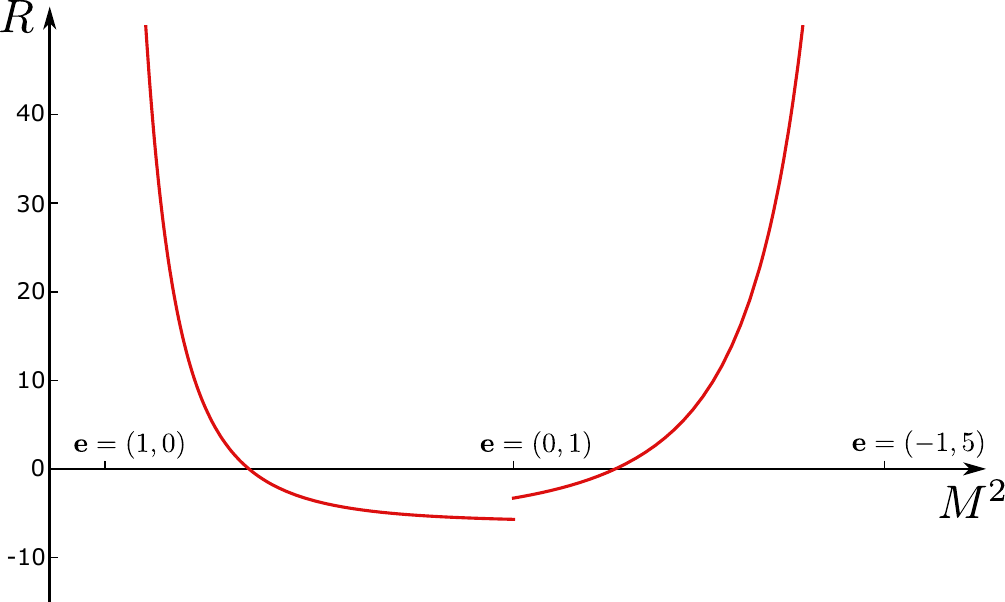}
\caption{Asymptotic behaviour of the classical scalar curvature $R^{\rm cl}_{\rm IIA}$ as a function of the M-theory modulus $M^2$ in the two-moduli example with a flop transition.}
\label{fig:R_ex2}
\end{center}
\end{figure}

We now consider a Calabi--Yau three-fold $X$ with two K\"ahler moduli, that was also studied in \cite{Gendler:2022ztv}. $X$ can be described as a hypersurface embedded in a toric fourfold constructed from a reflexive polytope and it admits a flop transition to a birationally equivalent Calabi--Yau $X'$. Given a basis of divisors $\{\mathcal{D}_1, \mathcal{D}_2\}$, we can expand the K\"ahler form as
\be
J = t^1 \omega_1 + t^2 \omega_2\, ,
\ee
and the (simplicial) extended K\"ahler cone, given by the union of the K\"ahler cones of the two phases, reads $\CK_{\rm ext}=\{(t^1,t^2)\ | \ 5t^1+t^2 \geq 0, t^2 \geq 0\}$. In the first phase $X$ the given divisor basis is Nef and the triple intersection numbers read
\be
\cK_{111}=5\, , \qquad \cK_{112}=5 \, , \qquad \cK_{122}=5\, , \qquad \cK_{222}=3\, ,
\ee
while in the phase $X'$ they become
\be     \label{Kabc_flop}
\cK_{111}=-15\, , \qquad \cK_{112}=5 \, , \qquad \cK_{122}=5\, , \qquad \cK_{222}=3\, .
\ee
One can perform the change of coordinates
\be
y^1=-t^1\, , \qquad y^2=5t^1+t^2\, ,
\ee
which amounts to take as basis of divisors $\mathcal{D}'_1=-\mathcal{D}_1+5\mathcal{D}_2$ and $\mathcal{D}'_2=\mathcal{D}_2$, and get the following triple intersection numbers
\be
\cK'_{111}=90\, , \qquad \cK'_{112}=30 \, , \qquad \cK'_{122}=10\, , \qquad \cK'_{222}=3\, .
\ee
The cone of effective divisors is given by
\be
\mathcal{E}=\{ (e^1,e^2)\in \mathbb{Z}^2 \ |\ e^1+e^2 \geq 0, 2e^1+e^2 \geq 0 \}\, ,
\ee
generated by $\mathcal{D}_A=\mathcal{D}_1-\mathcal{D}_2$ and $\mathcal{D}_B=-\mathcal{D}_1+2\mathcal{D}_2$. The M-theory moduli space is described by
\be
\begin{split}
\cM_{{\rm 5d}, X}^{\rm VM}  = \Bigl\{ (M^1,M^2) \ | \ M^i \geq 0 \quad  \text{and} \quad 5(M^1)^3 &+15(M^1)^2M^2 + 15M^1(M^2)^2 + 3(M^2)^3 =6 \Bigr\}\\
&\cup \\
\Bigl\{ (M^1,M^2) \ | \ M^1 \leq 0, 5M^1+M^2 \geq &\,\,0 \quad  \text{and} \\
-15(M^1)^3 +&15(M^1)^2M^2 + 15 M^1(M^2)^2 + 3(M^2)^3 =6 \Bigr\}
\end{split}
\ee
and the classical scalar curvature $R_{\rm IIA}^{\rm cl}$ along it is shown in figure \ref{fig:R_ex2}.

One can detect divergences at $(M^1,M^2)=\left(\left(\frac{6}{5}\right)^\frac{1}{3},0\right)$ and $(M^1,M^2)=\left(\frac{6}{110}\right)^\frac{1}{3}\left(-1,5\right)$, which respectively correspond to the limits $\bm{e}=(1,0)$ in the original variables $t^a$ and $\bm{e}'=(1,0)$ in the new variables $y^a$. On the other hand, the curvature reaches a negative constant value along the limits $\bm{e}=(0,1)$ and $\bm{e}'=(0,1)$, that correspond to approaching the facet where the flop transition occurs from the two sides. When going from one side to the other, the metric is continuous, but there is a finite jump in the scalar curvature, due to the change in the triple intersection numbers \eqref{Kabc_flop}. Let us analyse each elementary limit.
\begin{itemize}
\item Limit $t^1 \rightarrow \infty$ or $\bm{e}=(1,0)$: this is a $w=3$ limit with ${\bf k}=5$, the matrix
\begin{equation}
    {\bf K} = 
    \begin{pmatrix}
        5 & 5 \\
        5 & 5
    \end{pmatrix}
\end{equation}
has rank one and a kernel generated by $\bm{v}_{0}=(1,-1)$, that corresponds to the divisor class $\mathcal{D}_A$. Going to the new basis $\{\mathcal{D}_e,-\mathcal{D}_A \}$ we get
\be
\cK_{eee}=5\, , \qquad \cK_{AAA}=-2\, ,
\ee
with the rotated moduli $(t^e,t^A)=(t^1+t^2,t^2)$. In this limit the rigid prepotential reads
\be
\cF_{\rm rigid}^{\rm cl} = \frac{1}{3}\left(T^A\right)^3\, ,
\ee
where $T^A=T^2$ is the only dynamical field in the rigid theory. From this one can compute the metric $g_{A\bar{A}}=2t^2$ and the curvature
\be
R_{\rm rigid} =  \frac{1}{4(t^2)^3}\, ,
\label{Rrex2.1}
\ee
that multiplied by $4\mathcal{V}_X \sim \frac{10}{3}\phi^3$ reproduces the leading term of the type IIA classical moduli space curvature
\be
R_{\rm IIA}^{\rm cl} = \frac{5}{6(t_0^2)^3}\phi^3 + \mathcal{O}(\phi^2)\, .
\ee

\item Limit $t^2 \rightarrow \infty$ or $\bm{e}=(0,1)$: this is a $w=3$ limit where
\be
{\bf K} = 
\begin{pmatrix}
5 & 5 \\
5 & 3
\end{pmatrix},
\ee
has maximal rank. As a consequence, there is no rigid theory in the IR and the scalar curvature stays finite and it asymptotically tends to a negative constant, namely
\be     \label{Rfin1}
R_{\rm IIA}^{\rm cl} = -\frac{57}{10} + \mathcal{O}(\phi^{-1})\, ,
\ee
in agreement with the results of section \ref{s:limits}. However, as previously commented, this is not a real boundary of the M-theory moduli space, as crossing it gives rise to a flop transition and takes us to the second phase $X'$.

\item Limit $y^1 \rightarrow \infty$ or $\bm{e}'=(1,0)$: this is a $w=3$ limit with ${\bf k}'=90$, the matrix
\be
{\bf K}' = 
\begin{pmatrix}
90 & 30 \\
30 & 10
\end{pmatrix},
\ee
has rank one and a kernel generated by $\bm{v}'_{0}=(1,-3)$ -- or $\bm{v}_{0}=(-1,2)$ in the original coordinates -- that corresponds to the divisor class $\mathcal{D}_B$. In the new basis $\{\mathcal{D}_e,-\mathcal{D}_B \}$, we get
\be
\cK_{eee}=90\, , \qquad \cK_{BBB}=-9\, ,
\ee
with the rotated moduli $(t^e,t^B)=(y^1+\frac{1}{3}y^2,\frac{1}{3}y^2)$. One can then compute the rigid prepotential
\be
\cF_{\rm rigid}^{\rm cl} = \frac{3}{2}\left(T^B\right)^3\, ,
\ee
the metric $g_{B\bar{B}}=3y^2$ and finally the rigid scalar curvature
\be
R_{\rm rigid} = \frac{3}{2(y^2)^3} \, .
\label{Rrex2.2}
\ee
From this, taking into account that $\mathcal{V}_X \sim 15 \phi^3$, one can reproduce the scaling of
\be
R_{\rm IIA}^{\rm cl} = \frac{90}{(y_0^2)^3}\phi^3 + \mathcal{O}(\phi^2)\, .
\ee

\item Limit $y^2 \rightarrow \infty$ or $\bm{e}'=(0,1)$: this is a $w=3$ limit where
\be
{\bf K} = 
\begin{pmatrix}
30 & 10 \\
10 & 3
\end{pmatrix},
\ee
has maximal rank. It follows that there is no rigid theory in the IR and the scalar curvature tends to a negative constant, more precisely
\be
R_{\rm IIA}^{\rm cl} = -\frac{33}{10} + \mathcal{O}(\phi^{-1})\, .
\ee
As one can see, approaching the flop transition in the $X'$ phase gives a scalar curvature that differs by a finite constant from the curvature in the $X$ phase \eqref{Rfin1}.

\end{itemize}

\subsection*{Three-moduli example}

We consider the CY three-fold $\mathbb{P}^5_{1,1,2,8,12}[24]$, also studied in \cite{Klemm:1996bj,Lee:2019wij}, which can be seen as a K3 fibration over $\mathbb{P}^1_b$, where the fibre K3 is itself an elliptic fibration over $\mathbb{P}^1_f$. $X$ has three K\"ahler moduli and a simplicial K\"ahler cone generated by a basis of Nef divisors $\mathcal{D}_1,\mathcal{D}_2$ and $\mathcal{D}_3$. The K\"ahler form can be expanded as
\be
J=t^1 \omega_1 + t^2 \omega_2 + t^3 \omega_3\, ,
\ee
with $t^a \geq 0$, and the triple intersection numbers are
\be
\cK_{122}=2\, , \qquad \cK_{123}=1\, , \qquad \cK_{222}=8\, , \qquad \cK_{223}=4\, , \qquad \cK_{233}=2\, .
\ee
The cone of effective divisors is given by
\be
\mathcal{E} =\left\{(e^1,e^2,e^3)\in \mathbb{Z}^3 \ | \ e^2\geq 0, 2e^2+e^3 \geq 0, e^1+4e^2+2e^3 \geq 0 \right\}\, ,
\ee
and is generated by the divisor classes $\mathcal{D}_A=\mathcal{D}_1$, $\mathcal{D}_B=\mathcal{D}_2-2\mathcal{D}_3$ and $\mathcal{D}_C=-2\mathcal{D}_1+\mathcal{D}_3$, where geometrically $\mathcal{D}_A$ and $\mathcal{D}_B$ correspond to the K3 fibre and the Hirzebruch surface $\mathbb{F}_2$ that is the base of $X$ seen as an elliptic fibration, respectively. Here the 5d M-theory moduli space
\be
\begin{split}
\cM_{{\rm 5d}, X}^{\rm VM}  = \Bigl\{ (&M^1,M^2,M^3) \ | \ M^i \geq 0 \quad  \text{and} \\
&6M^1(M^2)^2 +6M^1M^2M^3 + 8(M^2)^3 + 12(M^2)^2M^3 + 6 M^2(M^3)^2 =6 \Bigr\}
\end{split}
\ee
is two-dimensional. By looking at the classical scalar curvature along this moduli space, we find that it asymptotes to a negative constant at the points $(M^1,M^2,M^3)=(\infty,0,0)$ and $(M^1,M^2,M^3)=(0,0,\infty)$, corresponding respectively to the limits $\bm{e}=(1,0,0)$ and $\bm{e}=(0,0,1)$, and the same happens for any positive linear combination of these vectors of charges. On the other hand, it diverges positively at the point $(M^1,M^2,M^3)=(0,\left(\frac{3}{4}\right)^{\frac{1}{3}},0)$, corresponding to the limit $\bm{e}=(0,1,0)$. Let us see more in detail the elementary limits in this case.
\begin{itemize}
\item Limit $t^1 \rightarrow \infty$ or $\bm{e}=(1,0,0)$: this is a $w=1$ limit, the matrix 
\be
{\bf K} = 
\begin{pmatrix}
0 & 0 & 0 \\
0 & 2 & 1 \\
0 & 1 & 0 
\end{pmatrix}
\ee
has rank two and a kernel generated by the  EFT string limit direction $\bm{e}$. As a consequence, there is no rigid IR theory and the type IIA classical scalar curvature goes like
\be
R_{\rm IIA}^{\rm cl} = -12 + \mathcal{O}(\phi^{-1})\, .
\ee

\item Limit $t^2 \rightarrow \infty$ or $\bm{e}=(0,1,0)$: this is a $w=3$ limit with ${\bf k}=8$, the matrix
\be
{\bf K} = 
\begin{pmatrix}
0 & 2 & 1 \\
2 & 8 & 4 \\
1 & 4 & 2 
\end{pmatrix},
\ee
has rank two and its kernel is generated by $\bm{v}_{0}=(0,1,-2)$, which corresponds to the divisor class $\mathcal{D}_B=\mathbb{F}_2$. To construct a basis, we follow \eqref{basisw3} and we pick the vector $\bm{v}_I=(-2,0,1)$, corresponding to the effective divisor class $\mathcal{D}_C$. In the new basis $\{\mathcal{D}_e,\mathcal{D}_C,-\mathcal{D}_B\}$ the triple intersection numbers become
\be
\cK_{eee}=8\, , \qquad \cK_{eCC}=-2\, , \qquad \cK_{CCB}=2\, , \qquad \cK_{BBB}=-8\, ,
\ee
and we have the rigid prepotential
\be
\cF_{\rm rigid}^{\rm cl} = \frac{4}{3} \left( T^B \right)^3 - (T_0^C)^2 T^B \, ,
\ee
where $T^B=\frac{1}{4}(T^1+2T^3)$ is the only dynamical field in the rigid theory. From here we compute the metric $g_{B\bar{B}}=8t^B$ and the curvature
\be
R_{\rm rigid} = \frac{4}{(t^1+2t^3)^3}\, ,
\ee
that multiplied by $4\mathcal{V}_X \sim \frac{16}{3}\phi^3$ reproduces
\be
R_{\rm IIA}^{\rm cl} = \frac{64}{3\left( t_0^1+2t_0^3 \right)^3}\phi^3 + \mathcal{O}(\phi^2)\, .
\ee

\item Limit $t^3 \rightarrow \infty$ or $\bm{e}=(0,0,1)$: this is a $w=2$ limit, the matrix
\be
{\bf K} = 
\begin{pmatrix}
0 & 1 & 0 \\
1 & 4 & 2 \\
0 & 2 & 0 
\end{pmatrix},
\ee
has rank two and the triple intersection numbers respect the structure of a smooth elliptic fibration \eqref{interfib}. The kernel is generated by $\bm{v}_0=(-2,0,1)$, namely the divisor class $\mathcal{D}_C$. To complete our new basis, see \eqref{basisw2}, we take the vector $\bm{v}_E=(0,1,0)$ that corresponds to ${\cal D}_E = {\cal D}_2$. In the basis $\{\mathcal{D}_2,\mathcal{D}_e,-\mathcal{D}_C\}$ the triple intersection numbers are
\be
\cK_{222}=8\, , \qquad \cK_{22e}=4\, , \qquad \cK_{2ee}=2\, , \qquad \cK_{2CC}=-2\, ,
\ee
and the rigid prepotential
\be
\cF_{\rm rigid}^{\rm cl} = T_0^2 \left( T^C \right)^2\, ,
\ee
from which we get $g_{C\bar{C}}=2t_0^2$ and $R_{\rm rigid}=0$, in agreement with an asymptotically constant classical curvature in the 4d EFT, see \eqref{Rneg},
\be
R_{\rm IIA}^{\rm cl} = -12 + \mathcal{O}(\phi^{-2})\, .
\ee
\end{itemize}

\subsection*{Three moduli with divergent emergent string limit}

Then we consider a CY that is a non-smooth elliptic fibration over a $\mathbb{P}^2$ base, which was studied in \cite{Hayashi:2023hqa}. This CY has three K\"ahler moduli and a simplicial K\"ahler cone. In a basis of Nef divisor we have the triple intersection numbers
\be
\begin{split}
&\cK_{122}=18\, , \qquad \cK_{123}=6\, , \qquad \cK_{133}=2\, ,\\
&\cK_{222}=63\, , \qquad \cK_{223}=15\, , \qquad \cK_{233}=3\, .
\end{split}
\ee
By looking at the behaviour of the classical moduli space scalar curvature along the 5d M-theory moduli space we observe the following. There is a cubic divergence at the point $(M^1,M^2,M^3)=(0,\left(\frac{2}{21}\right)^{\frac{1}{3}},0)$, corresponding to the limit $\bm{e}=(0,1,0)$, and a linear divergence at $(M^1,M^2,M^3)=(\infty,0,0)$, corresponding to $\bm{e}=(1,0,0)$. For positive linear combinations of these two vectors we get a cubic divergence as well. Finally, at $(M^1,M^2,M^3)=(0,0,\infty)$, namely $\bm{e}=(0,0,1)$, the curvature is a negative constant. Let us analyse these three elementary limits.
\begin{itemize}
\item Limit $t^1 \rightarrow \infty$ or $\bm{e}=(1,0,0)$: this is a $w=1$ limit, the matrix
\be
{\bf K} = 
\begin{pmatrix}
0 & 0 & 0 \\
0 & 18 & 6 \\
0 & 6 & 2 
\end{pmatrix},
\ee
has rank one and a null eigenvector besides $\bm{e}$, namely $\bm{v}_{0}=(0,1,-3)$, which corresponds to the divisor class $\mathcal{D}_A=\mathcal{D}_2-3\mathcal{D}_3$. To construct a basis we pick the vector $\bm{v}_I=(0,1,0)$, that represents the divisor class $\mathcal{D}_2$, see \eqref{eigenw1}. In the new basis $\{\mathcal{D}_e,\mathcal{D}_2,-\mathcal{D}_A\}$ the triple intersection numbers are
\be
\cK_{e22}=18\, , \qquad \cK_{222}=63\, , \qquad \cK_{22A}=-18\, , \qquad \cK_{AAA}=-9\, ,
\ee
and the rigid prepotential 
\begin{eqn}
    \cF_{\rm rigid}^{\rm cl} = \frac{3}{2}\left(T^{A}\right)^3+9\left(T^{2}_0\right)^2T^{A}\, ,
\end{eqn}
where $T^A=\frac{1}{3}T^3$ is the only dynamical field in the rigid theory. From here we compute the metric $g_{A\bar{A}}=9t^A$ and the curvature
\be
R_{\rm rigid} = \frac{3}{2(t^3)^3}\, ,
\ee
that multiplied by $4\mathcal{V}_X \sim 4(3t^2_0+t^3_0)^2\phi$ reproduces
\be
R_{\rm IIA}^{\rm cl} = \frac{6(3t^2_0+t^3_0)^2}{(t_0^3)^3} \phi + \mathcal{O}(\text{const})\, .
\ee

\item Limit $t^2 \rightarrow \infty$ or $\bm{e}=(0,1,0)$: this is a $w=3$ limit with ${\bf k}=63$. The matrix
\be
{\bf K} = 
\begin{pmatrix}
0 & 18 & 6 \\
18 & 63 & 15 \\
6 & 15 & 3 
\end{pmatrix},
\ee
has rank two and null eigenvector $\bm{v}_{0}=(-1,1,-3)$, which corresponds to the divisor class $\mathcal{D}_B=-\mathcal{D}_1+\mathcal{D}_2-3\mathcal{D}_3$. To construct a basis, we pick the vector $\bm{v}_I=(5,0,-6)$, representing the divisor class $\mathcal{D}_C=5\mathcal{D}_1-6\mathcal{D}_3$. In the new basis $\{\mathcal{D}_e,\mathcal{D}_C,-\mathcal{D}_B\}$ the triple intersection numbers are
\be
\begin{split}
\cK_{eee}=63\, , \qquad \cK_{eCC}=-252\, , \qquad \cK_{CCC}=1080\, ,\\
\cK_{CCB}=-36\, , \qquad \cK_{CBB}=18\, , \qquad \cK_{BBB}=-9\, ,
\end{split}
\ee
and the rigid prepotential 
\be
\cF_{\rm rigid}^{\rm cl} = \frac{3}{2}\left(T^{B}\right)^3-9T^{C}_0\left(T^{B}\right)^2+18\left(T^{C}_0\right)^2 T^{B}\, ,
\ee
where $T^B= \frac{1}{21} (6T^1 + 5T^3)$ is the only dynamical field in the rigid theory. From here we compute the metric $g_{B\bar{B}}=9t^B-18t_0^C$ and the curvature
\be
R_{\rm rigid} = \frac{3}{2(t^3)^3}\, ,
\ee
that multiplied by $4\mathcal{V}_X \sim 42\phi^3$ reproduces
\be
R_{\rm IIA}^{\rm cl} = \frac{63}{(t_0^3)^3} \phi^3 + \mathcal{O}(\phi^2)\, .
\ee

\item Limit $t^3 \rightarrow \infty$ or $\bm{e}=(0,0,1)$: this is a $w=2$ limit where the matrix
\be
{\bf K} = 
\begin{pmatrix}
0 & 6 & 2 \\
6 & 15 & 3 \\
2 & 3 & 0 
\end{pmatrix},
\ee
has maximal rank. As a consequence, there is no rigid IR theory and the type IIA classical scalar curvature goes like
\be
R_{\rm IIA}^{\rm cl} = -15 + \mathcal{O}(\phi^{-1})\, .
\ee
\end{itemize}

Following \eqref{basisw20} we take as a new basis the following vectors: $\bm{v}_{E}=(-1,1,0)$, that corresponds to the divisor class $\mathcal{D}_E=-\mathcal{D}_1+\mathcal{D}_2$ dual to the generic elliptic fiber, $\bm{v}_e= (0,0,1)$ and  $\bm{v}_I=(-3,2,0)$, namely the divisor class $\mathcal{D}_D = -3\mathcal{D}_1 + 2\mathcal{D}_2$. In this basis, the inverse gauge kinetic function has the leading behaviour
\be
\tilde{I}^{ab} =
\begin{pmatrix}
2(2t^1+3t^2) \phi^{-2} & -\frac{3t^2(4t^1+5t^2)}{2t^1+3t^2}\phi^{-2} & -2(t^1+t^2)\phi^{-2} \\
-\frac{3t^2(4t^1+5t^2)}{2t^1+3t^2}\phi^{-2} & \frac{1}{2t^1+3t^2} & -\frac{t^1+t^2}{2t^1+3t^2}\phi^{-1} \\
-2(t^1+t^2)\phi^{-2} & -\frac{t^1+t^2}{2t^1+3t^2}\phi^{-1} & \frac{1}{12}\phi^{-1} \\
\end{pmatrix} \quad + \text{ subleading terms}\, .
\ee
As one can clearly see, the weights for these three vectors are respectively, $w_{E}=2$, $w_e=0$ and $w_I=1$, in agreement with our  guess \eqref{ww20}.

\subsection*{Three-moduli non-smooth fibration}

Finally, we consider another non-smooth elliptic fibration over a $\mathbb{P}^2$ base, also studied in \cite{Hayashi:2023hqa}. It also has three K\"ahler moduli, a simplicial K\"ahler cone and in a Nef divisors basis we have
\be
\begin{split}
\cK_{111}=50\, , \qquad \cK_{112}=10\, , \qquad \cK_{122}=2\, , \qquad \cK_{113}=80\, , \qquad \cK_{123}=16\, ,\\ \cK_{223}=3\, , \qquad \cK_{133}=128\, , \qquad \cK_{233}=25\, , \qquad \cK_{333}=203\, .
\end{split}
\ee
In the M-theory moduli space the classical scalar curvature diverges positively at $(M^1,M^2,M^3)=(\left(\frac{3}{25}\right)^{\frac{1}{3}},0,0)$, corresponding to the limit $\bm{e}=(0,1,0)$, and at $(M^1,M^2,M^3)=(0,0,\left(\frac{6}{203}\right)^{\frac{1}{3}})$, corresponding to $\bm{e}=(0,0,1)$. Any positive linear combinations of these two vectors also gives a divergence. At the point $(M^1,M^2,M^3)=(0,\infty,0)$, reached along the limit $\bm{e}=(0,1,0)$, the curvature tends to a negative constant. Let us study the three elementary limits more in detail.
\begin{itemize}
\item Limit $t^1\rightarrow \infty$ or $\bm{e}=(1,0,0)$: this is a $w=3$ limit with ${\bf k}=50$. The matrix
\be
{\bf K} = 
\begin{pmatrix}
50 & 10 & 80 \\
10 & 2 & 16 \\
80 & 16 & 128 
\end{pmatrix}
\ee
has rank one and a kernel generated by the null eigenvectors $\bm{v}_{0,1}=(1,-5,0)$ and $\bm{v}_{0,2}=(8,0,-5)$, corresponding to the divisor classes $\mathcal{D}_A=\mathcal{D}_1-5\mathcal{D}_1$ and $\mathcal{D}_B=8\mathcal{D}_1-5\mathcal{D}_3$. In the basis $\{\mathcal{D}_e,-\mathcal{D}_A,-\mathcal{D}_B\}$ the triple intersection numbers become
\be
\cK_{eee}=50\, , \quad \cK_{AAA}=-50\, , \quad \cK_{AAB}=-25\, , \quad \cK_{ABB}=-75\, , \quad \cK_{BBB}=-225\, ,
\ee
and the rigid prepotential is
\be
\cF_{\rm rigid}^{\rm cl} = 25 \left[ \frac{1}{3}(T^A)^3 + \frac{1}{2} (T^A)^2 T^B + \frac{3}{2} T^A (T^B)^2 + \frac{3}{2} (T^B)^3 \right]\, ,
\ee
where this time we have two dynamical fields $T^A=\frac{1}{5}T^2$ and $T^B=\frac{1}{5}T^3$. From the last expression one can compute the metric
\be
g_{i\bar{j}}=25
\begin{pmatrix}
2t^A+t^B & t^A+3t^B\\
t^A+3t^B & 3t^A+9t^B
\end{pmatrix},
\ee
and then the rigid scalar curvature
\be
R_{\rm rigid} = \frac{3}{2}\left[\frac{1}{(t_0^2)^3}+\frac{5}{(t_0^2+3t_0^3)^3}\right]\, ,
\ee
and, taking into account that $\mathcal{V}_X \sim \frac{25}{3}\phi^3$, one can reproduce the 4d classical moduli space scalar curvature
\be
R_{\rm IIA}^{\rm cl} = 50 \left[\frac{1}{(t_0^2)^3}+\frac{5}{(t_0^2+3t_0^3)^3}\right] \phi^3 + \mathcal{O}(\phi^2)\, .
\ee

\item Limit $t^2\rightarrow \infty$ or $\bm{e}=(0,1,0)$: this is a $w=2$ limit where the matrix
\be
{\bf K} = 
\begin{pmatrix}
10 & 2 & 16 \\
2 & 0 & 3 \\
16 & 3 & 25 
\end{pmatrix}
\ee
has maximal rank. As a consequence, there is no rigid IR theory and the classical type IIA scalar curvature is
\be
R_{\rm IIA}^{\rm cl} = -15 + \mathcal{O}(\phi^{-1})\, .
\ee

We take as new basis the following vectors: $\bm{v}_{E}=(-1,0,1)$, that corresponds to the divisor class $\mathcal{D}_E=-\mathcal{D}_1+\mathcal{D}_3$ dual to the generic elliptic fiber, $\bm{v}_e= (0,1,0)$ and  $\bm{v}_I=(3,0,-2)$, namely the divisor class $\mathcal{D}_C = 3\mathcal{D}_1 - 2\mathcal{D}_3$. In this basis, the inverse gauge kinetic function has the leading behaviour

\be
\tilde{I}^{ab} =
\begin{pmatrix}
2(2t^1+3t^3) \phi^{-2} & -\frac{10(t^1)^2+32t^1 t^3+25(t^3)^2}{2t^1+3t^3}\phi^{-2} & -\frac{2(t^1+t^3)}{(t^2)^2}\phi^{-2} \\
-\frac{10(t^1)^2+32t^1 t^3+25(t^3)^2}{2t^1+3t^3}\phi^{-2} & \frac{1}{2t^1+3t^3} & -\frac{t^1+t^3}{2t^1+3t^3}\phi^{-1} \\
-\frac{2(t^1+t^3)}{(t^2)^2}\phi^{-2} & -\frac{t^1+t^3}{2t^1+3t^3}\phi^{-1} & \frac{1}{2}\phi^{-1} \\
\end{pmatrix}\, ,
\ee
up to subleading terms. As one can clearly see, the weights for these three vectors are, respectively, $w_{E}=2$, $w_e=0$ and $w_I=1$, again matching \eqref{ww20}.

\item Limit $t^3 \rightarrow \infty$ or $\bm{e}=(0,0,1)$: this is a $w=3$ limit with ${\bf k}=203$. The matrix
\be
{\bf K} = 
\begin{pmatrix}
80 & 16 & 128 \\
16 & 3 & 25 \\
128 & 25 & 203 
\end{pmatrix}
\ee
has rank 2 and its kernel is generated by $\bm{v}_{0}=(-1,-3,1)$, which corresponds to the divisor class $\mathcal{D}_{D}=-\mathcal{D}_1-3\mathcal{D}_2+\mathcal{D}_3$. To construct a basis, we pick the vector $\bm{v}_I=(25,-128,0)$, representing the divisor class $\mathcal{D}_F=25\mathcal{D}_1-128\mathcal{D}_2$. In the new basis $\{\mathcal{D}_e,\mathcal{D}_F,-\mathcal{D}_D\}$, the triple intersection numbers are
\be
\begin{split}
\cK_{eee}=203\, , \qquad \cK_{eFF}=-3248\, , \qquad \cK_{FFF}=838850\, ,\\
\cK_{FFD}=-16384\, , \qquad \cK_{FDD}=384\, , \qquad \cK_{DDD}=-9\, .
\end{split}
\ee
The rigid prepotential is
\be
\cF_{\rm rigid}^{\rm cl} = \frac{3}{2} (T^D)^3 - 192 (T^D)^2 T_0^D + 8192 T^D (T_0^F)^2\, ,
\ee
where the only dynamical field is $T^D=\frac{1}{203}(128T^1+25T^2)$. The rigid moduli space metric reads $g_{D\bar{D}}=9t^D-384t^F$ and the rigid scalar curvature
\be
R_{\rm rigid} = \frac{3}{2(t_0^2)^3}\, ,
\ee
which using $\mathcal{V}_X \sim \frac{203}{6}\phi^3$ correctly reproduces
\be
R_{\rm IIA}^{\rm cl} = \frac{203}{(t_0^2)^3}\phi^3 + \mathcal{O}(\phi^2)\, .
\ee
\end{itemize}


\section{Conclusions}
\label{s:conclu}

In this work we have analysed the asymptotic behaviour of the moduli space scalar curvature in a large class of trajectories of infinite distance. More precisely, we have focused on the vector multiplet sector of 4d $\CN=2$ supergravity theories obtained from compactifying type IIA on a Calabi--Yau three-fold $X$. Precisely in this setup, large-volume trajectories of infinite distance are known to provide counterexamples \cite{trenner2010asymptotic} to the expectation that the curvature $R_{\rm IIA}$ should be asymptotically negative \cite{Ooguri:2006in}. We have performed a general analysis of the asymptotic behaviour of $R_{\rm IIA}$, but instead of simply describing when Conjecture 3 of \cite{Ooguri:2006in} fails and when it does not, we have tried to understand the physics behind the different asymptotic behaviours of the curvature. Our results have led us to Conjecture \ref{conj:CC} which, if true, could lead us to a powerful criterion to describe the physics of infinite distance limits in terms of the scalar curvature in field space. 

Indeed, in the overall picture that we have obtained, an important role is played by those limits in which $R_{\rm IIA}$ diverges positively along the limit. In our setup, this only happens if one can see the infinite-distance trajectory as a gravity-decoupling limit in which a non-trivial field theory survives below the maximal cut-off $m_*$ set by the SDC leading tower. For the divergence to occur, the rigid field theory must be non-trivial in the sense that its curvature $R_{\rm rigid}$ does not vanish. Remarkably, we have obtained the same result for all the three classes of large-volume limits, although in a particular class ($w=2$ limits) world-sheet instanton corrections would be needed to generate $R_{\rm rigid}$. The different cases have been summarised in table \ref{tab:curv(w,rk)}, and the essential condition that determines if we have a rigid theory below the SDC scale is the rank $r$ of the matrix {\bf K} in \eqref{rank}. More precisely, the number of fields in the rigid theory is given by $n_V -r$, where $n_V = h^{1,1}(X)$ is the number of vector multiplets. Geometrically, this is the number of independent shrinkable, non-Nef divisors whose volume stays constant along the limit \eqref{limita}, and that host the rigid $\CN=2$ field theory that survives below the SDC scale. 

The existence of a rigid field theory below the SDC scale at any point along an infinite distance limit is interesting by itself, and it opens a new direction in the classification of infinite distance limits. Notice that the quantity that tells us whether such a rigid field theory should exist or not is the rank $r$ in \eqref{rank} which, as explained in section \ref{s:typeIIA}, is part of the classification of limits performed in \cite{Grimm:2018ohb,Grimm:2018cpv,Corvilain:2018lgw}. The details of the rigid theory, and in particular its curvature, seems to be additional information, which nevertheless is directly related to the asymptotic behaviour of the moduli space metric. It would then be very interesting to see if our findings can be generalised to the whole set of type II CY vector multiplet limits using the techniques of \cite{Grimm:2018ohb,Grimm:2018cpv,Corvilain:2018lgw}, and in particular to verify the content of Conjecture \ref{conj:CC}. In addition, it would be interesting to extend this analysis to the hypermultiplet moduli space of type II CY compactifications, although it could be that in this case the obstructions observed in \cite{Marchesano:2019ifh,Baume:2019sry} prevent divergences of the moduli space curvature at infinity. We hope to report on these topics in the future. 

Another result that is worth mentioning is the massive spectrum of states that comes with the rigid field theory. In our setup we have found that rigid field theories below the SDC scale come with their own infinite tower of charged states. These towers are different from the SDC leading tower and, even if they lie above the SDC cut-off $m_*$, they scale just like $m_*$ along the limit and they display an extremality factor that tends to infinity. This is particularly dramatic in $w=3$ limits where, due to the multiplicity of GV invariants within curves in a del Pezzo surface, their density is much larger than that of a Kaluza-Klein spectrum describing the tower of D0-branes. In our setup, the interpretation of $w=3$ in terms of M-theory compactified on $X$ relates this new tower of charged states with a decompactification to 5d with a non-Lagrangian SCFT at strong coupling \cite{Seiberg:1996bd,Morrison:1996xf,Douglas:1996xp,Intriligator:1997pq}. It would be interesting to see if this is a general lesson beyond our setup, and if there is a whole new casuistic to be explored. Additionally, it would be important to understand how these new towers influence the recent computation of the species scale \cite{Castellano:2021mmx,vandeHeisteeg:2022btw,vandeHeisteeg:2023ubh,vandeHeisteeg:2023dlw,Castellano:2023aum} and more precisely the role that it plays in recent tests of the Emergence Proposal \cite{Marchesano:2022axe,Castellano:2022bvr,Castellano:2023qhp,Blumenhagen:2023tev}.

From a more phenomenological perspective, it would be interesting to see if infinite distance limits with rigid field theories describe realistic corners of the string Landscape \cite{Ibanez:2012zz}. Several string theory scenarios use localisation properties of the compactification to engineer field theories that can be parametrically decoupled from gravity. In this way one can easily reproduce the disparity of strengths between the gravitational and gauge interactions observed empirically. Infinite distance limits with a rigid field theory below the SDC scale (or limits close to them) automatically implement this feature. Following the reasoning of section \ref{s:curvature} and assuming supersymmetry not far from the 4d cut-off scale this seems to indicate that, for EFTs that reproduce the observed gauge-gravitational hierarchy, the field space curvature in Planck units should be  positive. Reversing the logic, if the Curvature Criterion is correct, one could use the field space scalar curvature to scan over regions of the string Landscape, in order to select those corners that are more likely to reproduce our Universe. It would be  remarkable if something as simple as the field space curvature described one of the most mysterious hierarchies observed in Nature!

\vspace*{.15cm}

\centerline{\bf  Acknowledgments}


We would like to thank Alberto Castellano, Luis E. Ib\'a\~nez, Luca Martucci, Miguel Montero, Tom Rudelius, \'Angel M. Uranga, Timo Weigand and Max Wiesner for discussions.  This work is supported through the grants CEX2020-001007-S and PID2021-123017NB-I00, funded by MCIN/AEI/10.13039/501100011033 and by ERDF A way of making Europe. LM is supported by the fellowship LCF/BQ/DI21/11860035 and LP by the fellowship LCF/BQ/DI22/11940039 from ``La Caixa" Foundation (ID 100010434).


\appendix


\section{Asymptotic perturbative expansion of the metric}
\label{ap:details}

In this section we analyse the asymptotic scaling of the gauge kinetic function
\be     \label{Itilde}
\tilde{I}_{ab} = \frac{2}{3}\cK g_{ab} = \frac{3}{2\cK}\cK_a \cK_b - \cK_{ab}\, ,
\ee
along each class of limits. According to the type of limit, it might be convenient to rotate to a new basis.

\subsection*{$w=3$ limits}

We start by considering an EFT string limit with scaling index $w=3$. We can use the expansions \eqref{Kexpand} to compute \eqref{Itilde} and then go to the new basis $\{\bm{v}_e,\bm{v}_{I,p},\bm{v}_{0,\a}\}$, defined in \eqref{basisw3}.  Here and in the following the indices $p,q$ run over the basis elements of type $\bm{v}_{I}$, while $\a,\b$ are used for vectors of type $\bm{v}_{0}$. Taking into account that ${\bf K}_{ab} v_{0,\a}^b=0$, the rotated gauge kinetic function at leading order reads
\be     \label{Itilde_diag3}
\tilde{I}_{MN} = \tilde{I}_{ab} v^a_M v^b_N=
\begin{pmatrix}
\frac{1}{2}{\bf k}\phi & 2 {\bf k}_q^0 & \frac{3}{2} {\bf k}_\b^{00} \phi^{-1}\\
2 {\bf k}_p^0 & - {\bf K}_{pq} \phi & -{\bf K}_{\b p}^{0} \\
\frac{3}{2} {\bf k}_\a^{00} \phi^{-1} & -{\bf K}_{\a q}^{0} & - {\bf K}_{\a\b}^{0} 
\end{pmatrix} + \text{subleading terms} \, ,
\ee
where we have defined
\begin{align}
    {\bf k}^0_p \equiv {\bf K}_{ab} v_{I,p}^a t^b_0\, &, \quad {\bf k}^{00}_{\a} = \cK_{abc}v_{0,\a}^a t^b_0 t^c_0\, , \\\nonumber
    {\bf K}^{0}_{\a\b} = \cK_{abc} v_{0,\a}^a v_{0,\b}^b t^c_0 \,,\quad{\bf K}_{\a p}^{0}&\equiv \cK_{abc}  v_{I,p}^a v_{0,\a}^b t^c_0 \, , \quad {\bf K}_{pq}\equiv{\bf K}_{ab} v_{I,p}^a v_{I,q}^b
\end{align}
Notice that in \eqref{Itilde_diag3} the off-diagonal components might be further suppressed, even though we expect this to happen just for particular initial configurations of the saxions. In any case, this result is enough to compute the leading term of the inverse kinetic matrix
\be
\tilde{I}^{MN} =
\begin{pmatrix}
\frac{2}{\bf k}\phi^{-1} & \mathcal{O}(\phi^{-2}) & \mathcal{O}(\phi^{-2})\\
\mathcal{O}(\phi^{-2}) & - A^{pq} \phi^{-1} & \mathcal{O}(\phi^{-1})\\
\mathcal{O}(\phi^{-2}) & \mathcal{O}(\phi^{-1}) & -B^{\a\b}\\
\end{pmatrix} + \text{subleading terms} \, ,
\ee
where $A^{pq}$ and $B^{\a\b}$ are the inverses of ${\bf K}_{pq}$ and ${\bf K}_{\a\b}^{0} $, respectively. From here one can check that the weights of the vectors of the new basis indeed match with \eqref{ww3}.

\subsection*{$w=2$ limits, smooth elliptic fibration}

We now consider a $w=2$ limit, restricting to the case of a smooth elliptic fibration, where the triple intersection numbers take the form \eqref{interfib}. In this case we are already using a good basis, as one can see in \eqref{basisw2}, and actually there is no need to distinguish between $\bm{v}_0$ and $\bm{v}_e$ when computing the leading term of $\tilde{I}_{ab}$, as it is sufficient to split the coordinates into $\{t^E,t^\a\}$ to get
\be     \label{Itilde_diag2}
\tilde{I}_{ab} =
\begin{pmatrix}
\frac{\eta}{2 t^E} & \frac{1}{2} C_{\a\b} c_1^\a t^E  \\
\frac{1}{2} C_{\a\b} c_1^\b t^E & C_{\a\b} t^E
\end{pmatrix} + \text{subleading terms} \, ,
\ee
where we defined
\be
C_{\a\b} \equiv \left[ 2\frac{\eta_\a \eta_\b}{\eta} - \eta_{\a\b} \right]\, ,
\ee
and $\eta_\a=\eta_{\a\b}t^\b$ and $\eta=\eta_{\a\b}t^\a t^\b$. From \eqref{Itilde_diag2} one can compute the inverse gauge kinetic matrix
\be
\tilde{I}^{ab} =
\begin{pmatrix}
\frac{2 t^E}{(\eta e e)}\phi^{-2} & \mathcal{O}(\phi^{-2})  \\
\mathcal{O}(\phi^{-2}) & C^{\a\b} \frac{1}{t^E}
\end{pmatrix} + \text{subleading terms} \, ,
\ee
where $C^{\a\b}$ is the inverse of $C_{\a\b}$. From these expressions one can extract the weights \eqref{ww2}.

\subsection*{$w=1$ limits}

Finally, we consider a $w=1$ limit. We can use the expansions \eqref{Kexpand} to compute \eqref{Itilde} and then go to the new basis $\{\bm{v}_e,\bm{v}_{I,p},\bm{v}_{0,\a}\}$, defined in \eqref{eigenw1}. As before, the indices $p,q$ run over the basis elements of type $\bm{v}_{I}$, while $\a,\b$ are used for vectors of type $\bm{v}_{0}$. Taking into account that ${\bf K}_{ab} v_{0,\a}^b=0$, the rotated gauge kinetic function at leading order reads
\be     \label{Itilde_diag1}
\tilde{I}_{MN} =
\begin{pmatrix}
\frac{1}{2}{\bf k}^{00}\phi^{-1} & \left(\frac{ 1}{2}{\bf k}^{00}_q - \frac{ 1}{3}\frac{{\bf k}^{000}}{{\bf k}^{00}}{\bf k}^0_{q} \right)\phi^{-1} & \frac{1}{2} {\bf k}_\b^{00}\phi^{-1}\\
\left(\frac{ 1}{2}{\bf k}^{00}_p - \frac{ 1}{3}\frac{{\bf k}^{000}}{{\bf k}^{00}}{\bf k}^0_{p} \right)\phi^{-1} & D_{pq} \phi & \frac{{\bf k}^{00}_\b {\bf k}^{0}_p}{{\bf k}^{00}} - {\bf K}^0_{\b p} \\
\frac{1}{2} {\bf k}_\a^{00} \phi^{-1} & \frac{{\bf k}^{00}_\a {\bf k}^{0}_q}{{\bf k}^{00}} - {\bf K}^0_{\a q} & - {\bf K}_{\a\b}^{0} 
\end{pmatrix} 
\ee
up to subleading terms, where we have defined
\begin{align}
    {\bf k}^{00}\equiv {\bf K}_{ab}t^a_{0} t^b_{0}\,,\quad {\bf k}^{000}\equiv \mathcal{K}_{abc} t^a_{0} t^b_{0} t^c_{0} \,,\quad D_{pq}\equiv 2\frac{{\bf k}^{0}_p {\bf k}^{0}_q}{{\bf k}^{00}} - {\bf K}_{pq} \,,\quad {\bf k}^{00}_p\equiv\mathcal{K}_{abc} v_{I,p}^a t^b_{0} t^c_{0} \,.
\end{align}
The leading term of the inverse kinetic matrix
\be
\tilde{I}^{MN} =
\begin{pmatrix}
\frac{2}{{\bf k}^{00}}\phi & \mathcal{O}(\phi^{-1}) & \mathcal{O}(\phi^{-1})\\
\mathcal{O}(\phi^{-1}) &  D^{pq} \phi^{-1} & \mathcal{O}(\phi^{-1})\\
\mathcal{O}(\phi^{-1}) & \mathcal{O}(\phi^{-1}) & -B^{\a\b}\\
\end{pmatrix} + \text{subleading terms} \, ,
\ee
where $D^{pq}$ and $B^{\a\b}$ are the inverses of $D_{pq}$ and ${\bf K}_{\a\b}^{0}$, respectively. From these expressions, one can check that the weights of the vectors of the new basis indeed match with \eqref{ww1}.\\


\section{The scaling weight and degeneracy of EFT string limits}
\label{ap:degenerate}

In this section we explain the relations between the degeneracy of EFT string limits and their scaling weight $w$. From the viewpoint of \cite{Lanza:2021udy} an EFT string limit \eqref{limita} is said to be non-degenerate if the EFT string with charge $\bm{e}$ defining the limit is the only one becoming tensionless along the limit. If there are more EFT strings becoming tensionless, the limit is said to be degenerate and the order of degeneracy $p$ is the number of asymptotically tensionless independent (or equivalently elementary) EFT strings. In the following we will show that
\begin{enumerate}
\item $w=3 \qquad \qquad \qquad \qquad \quad \,\,\, \Rightarrow \quad p=n_V$, the limit is maximally degenerate.
\item $w=2$ (smooth  fibration) $ \quad \Rightarrow \quad p=n_V-1$, the limit is non-maximally degenerate.
\item $w=1 \qquad \qquad \qquad \qquad \quad \,\,\, \Rightarrow \quad p=1$, the limit is non-degenerate.
\end{enumerate}
Notice that for $w=2$ limits we are only considering smooth $T^2$ fibrations with the structure \eqref{interfib}. In the non-smooth case intermediate values of $p$ could be possible.

\begin{enumerate}
\item Consider a $w=3$ limit. Then ${\bf k}\neq 0$ and one has
\be
\cK = {\bf k} \phi^3 + \mathcal{O}(\phi^2)\, , \qquad \cK_a = {\bf k}_a \phi^2 + \mathcal{O}(\phi)\, ,
\ee
which imply, denoting by $\mathcal{T}_{a}/M_P^2 = \ell_a$ the tension of the $a^{\rm th}$ elementary EFT string,
\be
\frac{\mathcal{T}_a}{M_P^2} \sim \frac{3}{2} \frac{{\bf k}_a}{\bf k} \phi^{-1} \sim \phi^{-1}\, , \, \forall a\, .
\ee
Then, all the EFT strings become tensionless and we have a maximally degenerate limit, with $p=n_V$. Notice that in a $w=3$ limit, ${\bf k}_a = {\cal D}_a \cdot {\cal D}_{e} \cdot {\cal D}_e \neq 0, \, \forall a$, where ${\cal D}_e=e^a {\cal D}_a$ is the Nef divisor wrapped by the NS5 that defines the limit we are considering. One can show this by using the fact that two non-proportional Nef divisors always have non-trivial intersection and applying Lemma 5 of Appendix B of \cite{Lee:2019wij}.

\item Consider a $w=2$ that is a smooth $T^2$ fibration. Then one can split the charges as $(e^E, e^\a)$ and the limit in consideration is given by $\bm{e}=(0,e^\a)$. In this case one has ${\bf k}=0$, ${\bf k}_a=(\eta_{\a\b}e^\a e^\b,0)$ and
\be
\begin{split}
&\cK = 3 \eta_{\a\b}e^\a e^\b t^E_0 \phi^2 + \mathcal{O}(\phi)\, ,\\
\cK_E = \eta_{\a\b}e^\a e^\b &\phi^2 + \mathcal{O}(\phi) \, , \qquad \cK_\a = 2 \eta_{\a\b} e^\b t^E_0 \phi + \mathcal{O}(\text{const})\, .
\end{split}
\ee
Notice that $\eta_{\a\b}e^\b = {\cal D}_\a \cdot {\cal D}_e \neq 0, \, \forall \a$, where ${\cal D}_e=e^\b {\cal D}_\b$, because any two non-proportional Nef divisors have non-trivial intersection. This implies
\be
\frac{\mathcal{T}_a}{M_P^2} \sim
\begin{cases}
\phi^{-1} \quad a=\a\\
\text{const} \quad a=E
\end{cases}\, ,
\ee
and then the limit is degenerate with order $p=n_V-1$.

\item Consider a $w=1$ limit. Then ${\bf k} = {\bf k}_a = 0$ and one has
\be
\cK = 3 {\bf K}_{ab} t_0^a t_0^b \phi + \mathcal{O}(\text{const})\, , \qquad \cK_a = 2{\bf K}_{ab} t_0^b \phi + \mathcal{O}(\text{const})\, .
\ee
Notice that the fact that it is $w=1$ also implies that it is an elementary limit in the sense of footnote \ref{ft:elementary}, namely with just one non-vanishing charge.  As a consequence, ${\bf K}_{ab} t_0^b \neq 0$ if and only if $e^a=0$ and then
\be
\frac{\mathcal{T}_a}{M_P^2} \sim
\begin{cases}
\phi^{-1} \quad e^a \neq 0\\
\text{const} \quad e^a = 0
\end{cases}\, .
\ee
So the only EFT string becoming tensionless is the one with charge $e^a$ and the limit is non-degenerate.

\end{enumerate}



\bibliographystyle{JHEP2015}
\bibliography{papers}

\end{document}